\documentclass[fleqn,usenatbib]{mnras}

\usepackage{newtxtext,newtxmath}

\usepackage[T1]{fontenc}

\DeclareRobustCommand{\VAN}[3]{#2}
\let\VANthebibliography\thebibliography
\def\thebibliography{\DeclareRobustCommand{\VAN}[3]{##3}\VANthebibliography}

\usepackage{graphicx}	
\usepackage{amsmath}	

\usepackage{xcolor}
\usepackage{hyperref}
\usepackage{verbatim}
\usepackage{booktabs}
\usepackage{tabularx}
\usepackage[flushleft]{threeparttable}
\usepackage{caption}
\usepackage{color}
\usepackage{rotating}
\usepackage[normalem]{ulem}
 \useunder{\uline}{\ul}{}

\newcommand{\mearth}{M$_\oplus$}
\newcommand{\rearth}{R$_\oplus$}

\newcommand{\kms}{\ensuremath{\rm km\,s^{-1}}}
\newcommand{\ms}{\ensuremath{\rm m\,s^{-1}}}

\newcommand{\thisstar}{TOI-2134}

\newcommand{\rprstb}{0.03475}
\newcommand{\urprstb}{0.00038}

\newcommand{\inclb}{89.49}
\newcommand{\uinclb}{0.37}
\newcommand{\impb}{0.21}
\newcommand{\uimpb}{0.14}
\newcommand{\rplb}{2.69}
\newcommand{\urplb}{0.16}
\newcommand{\perplb}{9.2292005}
\newcommand{\uperplb}{0.0000063}
\newcommand{\ttransitb}{2459407.54493}
\newcommand{\uttransitb}{0.00027}
\newcommand{\impc}{0.464}
\newcommand{\uimpc}{0.042}
\newcommand{\rplc}{7.27}
\newcommand{\urplc}{0.42}
\newcommand{\ttransitc}{2459718.96939}
\newcommand{\uttransitc}{0.00020}
\newcommand{\rprstc}{0.09404}
\newcommand{\urprstc}{0.00078}

\newcommand{\offset}{2.30}
\newcommand{\offsetu}{+0.44}
\newcommand{\offsetd}{-0.45}
\newcommand{\Pb}{9.22923}
\newcommand{\Pbu}{+0.00004}
\newcommand{\Pbd}{-0.00004}
\newcommand{\Kb}{3.40}
\newcommand{\Kbu}{+0.28}
\newcommand{\Kbd}{-0.29}
\newcommand{\Ckb}{0.21}
\newcommand{\Ckbu}{+0.06}
\newcommand{\Ckbd}{-0.08}
\newcommand{\Skb}{-0.07}
\newcommand{\Skbu}{+0.06}
\newcommand{\Skbd}{-0.09}
\newcommand{\tperib}{2459407.89}
\newcommand{\tperibu}{+0.45}
\newcommand{\tperibd}{-0.49}

\newcommand{\Pc}{95.50}
\newcommand{\Pcu}{+0.36}
\newcommand{\Pcd}{-0.25}
\newcommand{\Kc}{9.74}
\newcommand{\Kcu}{+1.60}
\newcommand{\Kcd}{-1.63}
\newcommand{\Ckc}{0.59}
\newcommand{\Ckcu}{+0.13}
\newcommand{\Ckcd}{-0.16}
\newcommand{\Skc}{-0.57}
\newcommand{\Skcu}{+0.19}
\newcommand{\Skcd}{-0.13}
\newcommand{\tperic}{2459721.20}
\newcommand{\tpericu}{+1.52}
\newcommand{\tpericd}{-1.23}

\newcommand{\eccb}{0.06}
\newcommand{\eccbu}{+0.03}
\newcommand{\eccbd}{-0.04}
\newcommand{\omegab}{1.91}
\newcommand{\omegabu}{+0.32}
\newcommand{\omegabd}{-0.34}

\newcommand{\omegac}{02.32}
\newcommand{\omegacu}{+0.22}
\newcommand{\omegacd}{-0.32}
\newcommand{\eccc}{0.67}
\newcommand{\ecccu}{+0.05}
\newcommand{\ecccd}{-0.06}

\newcommand{\Mb}{9.13}
\newcommand{\Mbu}{+0.78}
\newcommand{\Mbd}{-0.76}
\newcommand{\Mc}{41.89}
\newcommand{\Mcu}{+7.69}
\newcommand{\Mcd}{-7.83}

\newcommand{\Pk}{45.78}
\newcommand{\Pku}{+5.56}
\newcommand{\Pkd}{-5.31}
\newcommand{\Kk}{5.52}
\newcommand{\Kku}{+0.67}
\newcommand{\Kkd}{-0.71}
\newcommand{\Harmk}{0.48}
\newcommand{\Harmku}{+0.05}
\newcommand{\Harmkd}{-0.05}
\newcommand{\Evok}{25.05}
\newcommand{\Evoku}{+8.48}
\newcommand{\Evokd}{-8.53}
\newcommand{\jit}{0.91}
\newcommand{\jitu}{+0.14}
\newcommand{\jitd}{-0.13}

\title[TOI-2134 system]{A hot mini-Neptune and a temperate, highly eccentric sub-Saturn around the bright K-dwarf TOI-2134
\thanks{Based on observations made with the Italian Telescopio Nazionale Galileo (TNG) operated on the island of La Palma by the Fundación Galileo Galilei of INAF (Istituto Nazionale di Astrofisica) at the Spanish Observatorio del Roque de los Muchachos of the Instituto de Astrofísica de Canarias}}

\author[F. Rescigno et al.]
{F. Rescigno$^{1}$, 
G. Hébrard$^{2,3}$, 
A. Vanderburg$^{4}$, 
A. W. Mann$^{5}$,  
A. Mortier$^{6}$, 
S. Morrell$^{1}$, 
L. A. Buchhave$^{7}$, 
\newauthor
K. A. Collins$^{8}$, 
C. R. Mann$^{9,10}$, 
C. Hellier$^{11}$,  
R. D. Haywood$^{1}$\footnote{STFC Ernest Rutherford Fellow}, 
R. West$^{12}$,  
M. Stalport$^{13, 14}$, 
N. Heidari$^{2}$, 
\newauthor
D. Anderson$^{12}$, 
C. X. Huang$^{15}$, 
M. L\'opez-Morales$^{8}$, 
P. Cort\'es-Zuleta$^{16}$, 
H. M. Lewis$^{17}$, 
X. Dumusque$^{14}$, 
\newauthor
I. Boisse$^{18}$, 
P. Rowden$^{19}$, 
A. Collier Cameron$^{20,21}$,  
M. Deleuil$^{16}$, 
M. Vezie$^{4}$, 
F. A. Pepe$^{14}$,  
X. Delfosse$^{22}$, 
\newauthor
D. Charbonneau$^{8}$, 
K. Rice$^{23,24}$, 
O. Demangeon$^{25}$, 
S. N. Quinn$^{8}$, 
S. Udry$^{14}$, 
T. Forveille$^{22}$, 
J. N. Winn$^{26}$, 
\newauthor
A. Sozzetti$^{27}$, 
S. Hoyer$^{16}$, 
S. Seager$^{4,28,29}$, 
T. G. Wilson$^{20,21}$, 
S. Dalal$^{1}$, 
E. Martioli$^{2,30}$, 
S. Striegel$^{31}$, 
\newauthor
W. Boschin$^{32,33,34}$, 
D. Dragomir$^{35}$, 
A. F. Martínez Fiorenzano$^{32}$, 
R. Cosentino$^{32}$, 
A. Ghedina$^{32}$, 
\newauthor
L. Malavolta$^{36,37}$, 
L. Affer$^{38}$, 
B. S. Lakeland$^{1}$, 
B. A. Nicholson$^{39,40}$ 
S. Foschino$^{41}$, 
A. W\"{u}nsche$^{41}$, 
\newauthor
K. Barkaoui$^{42,28,33}$, 
G. Srdoc$^{43}$, 
J. Randolph$^{44}$, 
B. Guillet$^{44}$, 
D. M. Conti$^{44}$, 
M. Ghachoui$^{42,45}$, 
\newauthor
M. Gillon$^{42}$, 
Z. Benkhaldoun$^{45}$, 
F. J. Pozuelos$^{42,46,13}$, 
M. Timmermans$^{42}$, 
E. Girardin$^{47}$, 
S. Matutano$^{48}$, 
\newauthor
P. Bosch-Cabot$^{48}$, 
J. A. Mu\~noz$^{49,50}$, 
R. For\'es-Toribio$^{49,50}$ 
\\
\\
\emph{Affiliations are listed at the end of the paper}}

\date{Accepted 2023 November 19. Received 2023 November 20; in original form 2023 May 5.}

\pubyear{2023}

\begin{document}
\label{firstpage}
\pagerange{\pageref{firstpage}--\pageref{lastpage}}
\maketitle

\begin{abstract}
We present the characterisation of an inner mini-Neptune in a \perplb$\pm$\uperplb \,day orbit and an outer mono-transiting sub-Saturn planet in a \Pc$^{\Pcu}_{\Pcd}$ day orbit around the moderately active, bright ($m_{v}$ = 8.9 mag) K5V star TOI-2134. Based on our analysis of five sectors of TESS data, we determine the radii of TOI-2134b and c to be 2.69$\pm$0.16 R$_{\oplus}$ for the inner planet and \mbox{7.27$\pm$0.42 R$_{\oplus}$} for the outer one. We acquired 111 radial-velocity spectra with HARPS-N and 108 radial-velocity spectra with SOPHIE. After careful periodogram analysis, we derive masses for both planets via Gaussian Process regression: \Mb$^{\Mbu}_{\Mbd}$ M$_{\oplus}$ for TOI-2134b and \Mc$^{\Mcu}_{\Mcd}$ M$_{\oplus}$ for TOI-2134c. We analysed the photometric and radial-velocity data first separately, then jointly. The inner planet is a mini-Neptune with density consistent with either a water-world or a rocky core planet with a low-mass H/He envelope. The outer planet has a bulk density similar to Saturn's. The outer planet is derived to have a significant eccentricity of \eccc$^{\ecccu}_{\ecccd}$ from a combination of photometry and RVs. We compute the irradiation of TOI-2134c as 1.45$\pm$0.10 times the bolometric flux received by Earth, positioning it for part of its orbit in the habitable zone of its system. We recommend further RV observations to fully constrain the orbit of TOI-2134c. With an expected Rossiter-McLaughlin (RM) effect amplitude of \mbox{7.2$\pm$1.3 \ms}, we recommend \mbox{TOI-2134c} for follow-up RM analysis to study the spin-orbit architecture of the system. We calculate the Transmission Spectroscopy Metric, and both planets are suitable for bright-mode NIRCam atmospheric characterisation.
\end{abstract}




\section{Introduction}
\label{sec:intro}

Since the discovery of the first exoplanet circa 30 years ago, more than 5000 have been detected and confirmed. Radial-velocity surveys performed with instruments such as the High Accuracy Radial-velocity Planet Searcher (HARPS) coupled with the Kepler photometric mission started discovering a sub-population of small exoplanets in short (under 100 days) orbits \citep{Mayor2008, Lovis2009, Fressin2009, Borucki2011,Batalha2013}.
Given their abundance in our galaxy \citep{Chabrier2000, Winters2015}, and their low mass and size, K and M dwarf stars are prime candidates for small-exoplanet searches and demographic-focused studies \citep{Dressing2013, Crossfield2015, Astudillo-Defru2017, Pinamonti2018, West2019, Rice2019, Burt2020}.\\

The transition point between rocky super-Earths and gaseous Neptunes is still debated \citep{Fulton2017, Luque2021}. \citet{Otegi2020} shows that this transition range is between \mbox{5-25 M$_{\oplus}$} and 2-3 R$_{\oplus}$, but several factors play into the composition of these planets. Some studies report that all planets under 1.6 R$_{\oplus}$ must be rocky \citep{Rogers2015, LopezMorales2016}. Others give more importance to the effects of irradiation: less irradiated planets are more likely to maintain a gaseous envelope, while more irradiated ones are typically rocky \citep{Hadden2014, Jontof-Hutter2016}. \cite{Owen2019} explores how planetary magnetic fields can also decrease their mass-loss rates and therefore alter the composition of the planetary cores.
A continuous effort in the detection of small planets, and in the precise characterisation of their masses and sizes is therefore vital to reach a consensus on which parameters affect planetary composition.\\

On the other hand, our understanding of long-period planets is also lacking. The great majority of transit-detected exoplanets have periods shorter than 75 days \citep{Jiang2019}. Longer-period planets are harder to detect and determining their masses can be challenging. Moreover, the baselines of most photometric surveys also limit their detection. This "missing" population hampers studies of planet demographics, of planet formation, and of how planetary characteristics depend on the host star \citep{Winn2011, Johnson2010}.

Temperate giants are located in a period valley, between 10 and 100 days, where gas planets are less frequent \citep{Udry2003,Wittenmyer2010}.
Although more challenging to study, these cooler planets are valuable sources of information. For starters, temperate giant planets represent the middle step between the short-period Hot Jupiters and the gas giants of our own solar system. They therefore can serve as bridges between their respective formation and migration theories \citep{Huang2016}. The composition of giant planets depends not only on the composition of the protoplanetary disk, but also on their location at birth and migration history. Consequently, studying their metal enrichment levels can constrain the processes driving core formation and envelope enrichment \citep{Thorngren2016, Mordasini2016}.
Recent studies have also shown that long-period planets are correlated to and influence the dynamical evolution of the short-period planets within their systems \citep{Zu2018,Bryan2019}. Moreover, theoretical models predict that the formation of inner Earth-like planets is significantly dependent on the presence of quickly-accreted cold giants \citep{Morbidelli2022}.
Due to their lower effective temperatures, the atmospheres of temperate giants produce entirely different molecular abundances and potentially can contain disequilibrium chemistry by-products \citep{Fortney2020}, making long-period gas planets valuable targets for atmospheric characterisation. Their atmospheres are less affected by temperature-induced inflation, which in turn allows us to use cooling models of planet evolution  to constrain atmospheric metallicity \citep{Ulmer-Moll2022}.
Additionally, there is a clear split in the eccentricity distribution of long-period planets. They are divided into a first group of objects with significantly high eccentricities and a second group with consistently nearly circular orbits \citep{Petrovich2016}. No clear cause of this bimodality has been found yet.\\

The numerous and highly varied scientific interests in exoplanet detection and characterisation have in the years motivated many space-based missions and ground-based instruments, including the second-generation HARPS-N \citep{Cosentino2012} and the SOPHIE \citep{Perruchot2008} spectrographs.
Paired with space photometric missions \citep[e.g., ][]{Ricker2015}, the combination of transit photometry and radial velocity (RV) makes the determination of precise planetary masses and radii possible. The precision of RV surveys has been steadily improving and the current uncertainty level reaches down to the tens of centimetres per second \citep{Jurgenson2016, Thompson2016, Pepe2021}, but the biggest obstacle remains stellar variability \citep{Fischer2016, Crass2021}.
Great care is required when accounting for and modelling stellar activity in order to obtain accurate orbital solutions and to accurately and precisely determine planetary masses. To do so Gaussian Process (GP) regression coupled with Monte Carlo Markov Chain parameter space exploration has been implemented in this paper and its specifics will be discussed in Section \ref{sec:rv}.\\

In this paper we characterise the high proper-motion, bright ($m_v$=8.9 mag) K5-dwarf TOI-2134 and its planetary system. We detect a multi-transiting mini-Neptune in a short circular orbit and an outer temperate sub-Saturn planet. 
We also propose these targets for Rossiter-McLaughlin effect \citep{Rossiter1924, McLaughlin1924, Queloz2000} follow-up and for atmospheric characterisation.

This paper is structured as follows: in Section \ref{sec:data} we describe the photometric and spectroscopic data used in our analysis of the system.
In Section \ref{sec:star} we characterise the host star with four independent techniques. In Section \ref{sec:activity} we include the analysis of the stellar signals and its activity proxies to identify the stellar rotational period. In Section \ref{sec:transit} and \ref{sec:rv} we fit the photometric data for transit parameters and perform a GP regression on the radial-velocity data to determine the planets' masses, radii and orbit characteristics. Results can be found in Tables \ref{tab:phot} and \ref{tab:separategp}. In Section \ref{sec:joint} we combine the two datasets and perform a joint photometric and RV analysis, with reults in Table \ref{tab:joint}. Final results are shown in Table \ref{tab:big} and addressed in Section \ref{sec:results}, together with proposed follow-ups.

\begin{figure*}
	\includegraphics[width=16cm]{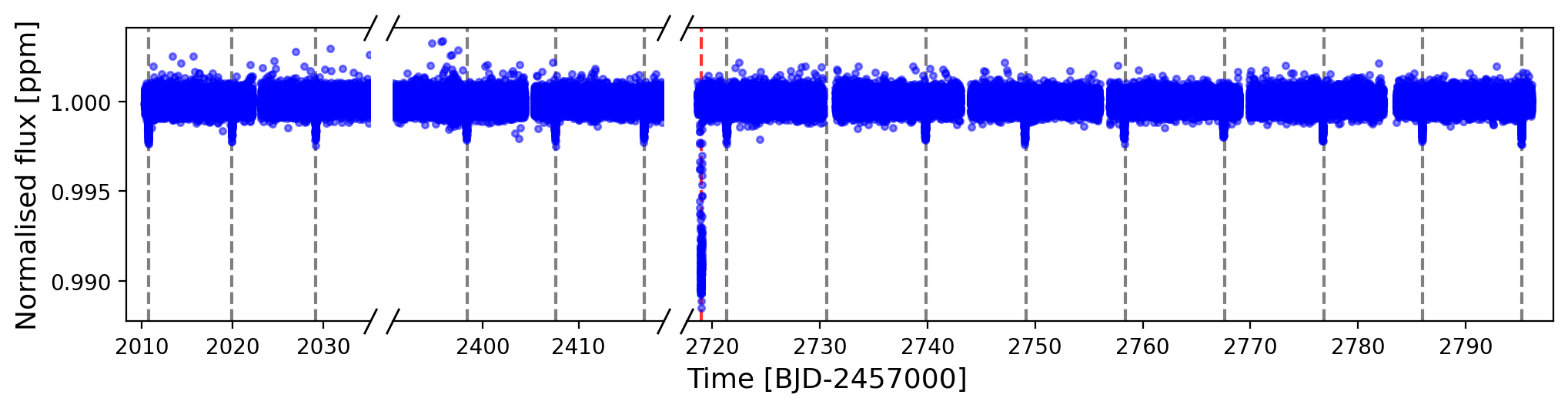}
    \caption{TESS normalised lightcurve over 5 sectors. 14 transits of an inner planet and a mono-transit of an outer planet can be seen and are indicated by the grey and red dashed lines respectively.}
    \label{fig:TESS_data}
\end{figure*}
\begin{figure*}
    \centering
	\includegraphics[width=15cm]{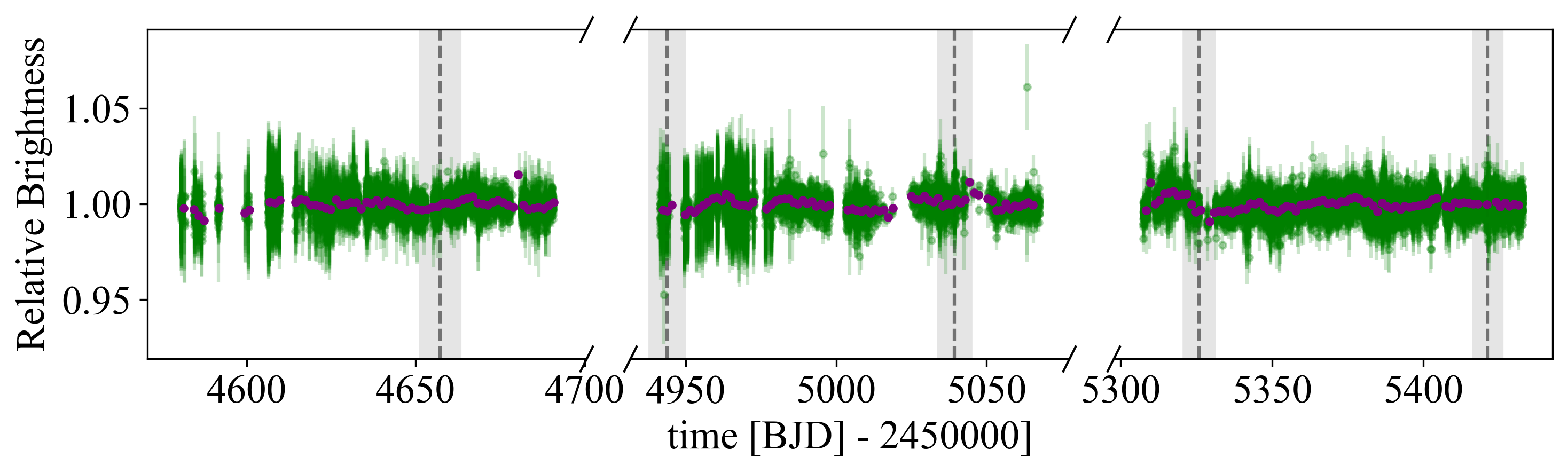}
    \caption{WASP normalised flux against Julian Date over the three years of coverage. All datapoints are plotted in green with errorbars, daily averages are plotted in purple. The predicted transits of TOI-2134c are plotted as grey dashed lines, while their uncertainties are plotted as grey shaded areas. As we address in Section \ref{sec:results}, we did not detect any transit.}
    \label{fig:WASP_data}
\end{figure*}

\section{Data}
\label{sec:data}
\subsection{TESS Photometry}
\label{sec:TESS}
TOI-2134, also known as TIC 75878355 in the TESS Input Catalog \citep{Stassun2018}, was observed by NASA's Transiting Exoplanet Survey Satellite (TESS: \citealt{Ricker2015}) mission in 2-minute cadence mode over five sectors (Sectors 26, 40, 52, 53 and 54) for a total of 88431 datapoints between BJD 2459010 and 2459035 (2020 June 9 to July 4), BJD 2459390 and 2459418 (2021 June 24 to July 22), and BJD 2459718 and 2459797 (2022 May 18 to August 5). The data were originally processed by the TESS Science Processing Operation Centre (SPOC) pipeline based at NASA Ames Research Center \citep{Jenkins2016}. However, Sector 40 showed strong residual systematics after the SPOC correction, so we performed our own systematics corrections of the SPOC Simple Aperture Photometry (SAP) light curves \citep{Twicken2010,Morris2020}. In particular, we modelled the systematics as a sum of moments of the spacecraft quaternion time series (e.g. \citealt{Vanderburg2019}) and modelled long-term variations with a basis spline. We also included a term for variations in the background flux in our model. We performed the model fit using an analytic linear least squares fit, excluding transits and iterating the fit several times to remove outliers. The resulting light curve was similar to the SPOC light curve (with slightly lower scatter) in most sectors, and yielded a major improvement in the problematic sector 40.

The transit signature of a TOI-2134b candidate was initially identified in a transit search conducted by the SPOC of Sector 26 on 24th July 2020 with an adaptive, noise-compensating matched filter \citep{Jenkins2002, Jenkins2010}. Diagnostic tests were also conducted to help make or break the planetary nature of the signal \citep{Twicken2010}.
The transit signatures for the TOI-2134b candidate  were also detected in a search of Full Frame Image (FFI) data by the Quick Look Pipeline (QLP) at MIT \citep{Huang2020a,Huang2020b} for Sector 40. A larger transit was detected by both QLP and the SPOC in searches including Sector 52. This transit was attributed to a second planetary candidate in the system, TOI-2134c. It appears to be a mono-transit and it did not re-occur in the following 75 days.
The TESS Science Office (TSO) reviewed the vetting information and issued an alert on 7th August 2020 for TOI-2134b and on 28th July 2022 for TOI-2134c \citep{Guerrero2021}. The signal for the candidate TOI-2134b was repeatedly recovered as additional observations were made in sectors 26, 40, 52, 53, and 54, and the transit signatures passed all the diagnostic tests presented in the Data Validation reports. The difference image centroiding figure and difference images for the multi-sector Sector 26 - Sector 55 run for candidate TOI-2134b show that the centroid of the transit source is consistent with the target star of interest. The host star is located within 3.2$\pm$3.7 arcsec of the source of the transit signal for candidate TOI-2134b and within 0.98$\pm$2.59 arcsec of the source of the transit signal for candidate TOI-2134c.
We flattened the light curve by simultaneously fitting transit models for the two planets along with a basis spline to model long-term variations, and then subtracting the long-term variations (a strategy similar to \citealt{Vanderburg2016}, except without a simultaneous systematics model; see also \citealt{Pepper20}).
The systematics-corrected and flattened TESS data are shown in Fig. \ref{fig:TESS_data}.
To better constrain the characteristics of the mono-transiting long-period planet candidate, we launched a ground- and space-based photometric observing campaign to catch a second transit.

\subsection{LCOGT Photometry}
\label{sec:LCO}
The Las Cumbres Observatory Global Telescope (LCOGT: \citealt{Brown2013}) network observed the star between BJD 2459808 and 2459818 (2022 August 17 to 27), when preliminary ephemeris prediction suggested the outer planet would re-transit.

Due to an unfortunate combination of bad weather and low visibility, only a possible egress was detected. However, the LCO 0m4 SBIG detectors are very susceptible to strong systematics and several combinations of comparison stars and aperture sizes need to be examined to assess the overall reliability of a light curve feature, especially for ingress- or egress-only events. When using a different choice of comparison stars, a convincing egress was no longer present in the data. The apparent egress was, in fact, proven to be highly dependent on the choice of comparison star set. For this reason, we could not claim this egress as a detected transit on its own and we do not include this data in our analysis.

We also attempted a TRansiting Planets and PlanetesImals Small Telescope (TRAPPIST) North \citep{Barkaoui2017} observation of the outer plant on 22nd August 2022, but it was unsuccessful.

\subsection{NEOSSat Photometry}
\label{sec:NEOSSat}
The position in the sky of TOI-2134 is such that it is not observable after late-October, which precluded the chance of a second ground-based campaign to detect a third transit of the outer planet candidate since the TESS detection. We therefore turned to space observations. TOI-2134 is outside of the CHEOPS field of view, but it is visible to the agile space telescope Near Earth Object Surveillance Satellite (NEOSSat: \citealt{Hildebrand2004, Fox2022}). NEOSSat is a Canadian microsatellite orbiting the Earth in a Sun-synchronous orbit of approximately 100 minutes. It was originally deployed to study near-Earth satellites, but it also performs well for follow-up observations of large exoplanets transiting bright stars. It carries a 15cm f/6 telescope, with spectral range between 350 and 1050 nm and a field of view of 0.86$\times$0.86 degrees.

NEOSSat observed TOI-2134 unevenly between BJD 2459898 and 2459910 (2022 November 14 to 26) with a 70s cadence for a total of 3364 datapoints. Multiple sets of observations through the run show significant unpredictable offsets that are usually corrected with calibration on reference stars. In these orbits, however, the reference stars behave differently from each other and the correction is less precise. This is probably due to image artefacts, as the detector and readout process have quite noticeable imperfections. These high-variance orbits have been flagged in the dataset and appear often enough to prevent a clear confirmation of a transit.\\

\begin{figure*}
    \includegraphics[width=15cm]{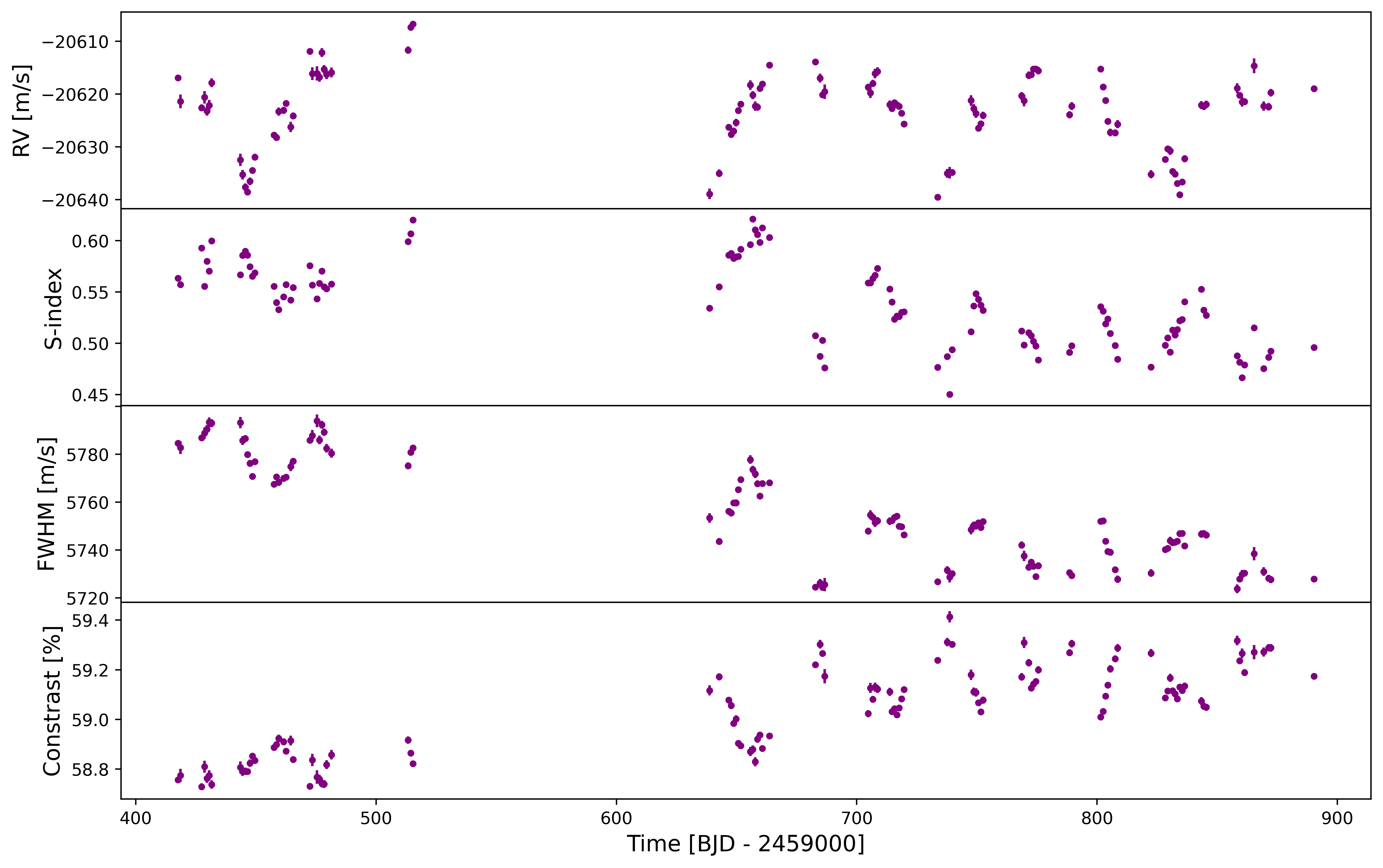}
    \hspace{-0.5cm}
	\includegraphics[width=15cm]{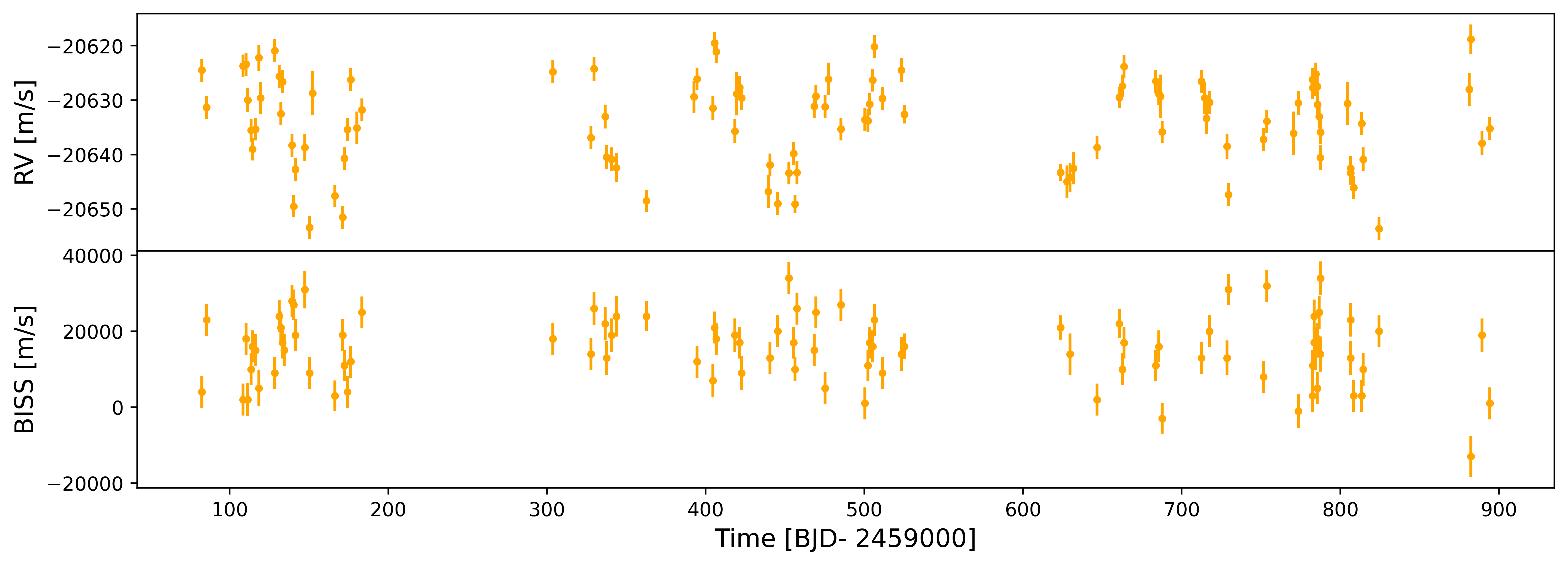}
    \caption{Plots of the HARPS-N and SOPHIE radial-velocity data alongside the chosen activity proxies for each dataset (see Section \ref{sec:act,rv}). From the top: HARPS-N RVs, S-index, full width at half maximum (FWHM) and contrast in purple, followed by SOPHIE RVs and their bisector span (BISS) in orange. Notice the different time axes. All error bars are plotted, but some are too small to be clearly visible in HARPS-N data. We plot only the activity proxies used in the later analysis (for more information see Section \ref{sec:act,rv}).}
    \label{fig:all_rv}
\end{figure*}
\subsection{WASP Photometry}
\label{sec:WASP}
TOI-2134 was also observed over 3 years by the Wide Angle Search for Planets (WASP: \citealt{Pollacco2006, Wilson2008}) with coverage of about 120 nights per year. The data cover similar three-month spans between BJD 2454580 to 2454690 (2008 April 23 to August 11), BJD 2454941 to 2455067 (2009 April 19 to August 23) and BJD 2455307 to 2455432 (2010 April 20 to August 23).
A total of 23097 datapoints were obtained and reduced with the SuperWASP pipeline \citep{Pollacco2006}. No planetary transit was detected. However, the long baseline, over three years long, allows for long-term monitoring of the stellar activity and of the rotational period of the host star, as shown in Section \ref{sec:act,photo}. All data are shown in Fig. \ref{fig:WASP_data}.

\subsection{HARPS-N Spectroscopy}
\label{sec:harpsn} 

We collected a total of 111 radial-velocity observations of TOI-2134 over two seasons with the High Accuracy Radial-velocity Planet Searcher for the Northern hemisphere spectrograph (\mbox{HARPS-N}: \citealt{Cosentino2012, Cosentino2014}) installed on the 3.6m Telescopio Nazionale Galileo (TNG) at the Observatorio del Roque de Los Muchachos in La Palma, Spain. HARPS-N is an updated version of HARPS at the ESO 3.6-m \citep{Mayor2003}. The spectrograph covers the wavelength range of 383-691 nm, with an average resolution $R$ = 115,000.
The first 32 spectra were collected between BJD 2459417 and 2459515 (2021 July 21 to October 27), and the next 79 were collected between BJD 2459638 and 2459890 (2022 February 27 to November 6). All data were observed under the Guaranteed Time Observations (GTO) programme with the standard observing approach of one observation per night.
The average exposure time for TOI-2134 was 900s with an average signal-to-noise ration (SNR) at \mbox{550 nm} of $\sim$100. 
RVs and activity indicators were extracted using the 2.3.5 version of the Data Reduction Software (DRS) adapted from the ESPRESSO pipeline (see \citealt{Dumusque2021}) and computed using a K6-type numerical weighted mask. The radial-velocity data show a peak-to-peak dispersion of 35 \ms, with standard rms of 7.3 \ms and mean uncertainty of 0.7 \ms.

Several proxies are extracted by the standard DRS pipeline, including (but not limited to) the full width at half maximum (FWHM) and the contrast of the cross-correlation function (CCF), and the S-index. The mentioned data are plotted in purple in Fig. \ref{fig:all_rv}.
The reasoning behind the selection of plotted proxies is addressed in Section \ref{sec:act,rv}.

\subsection{SOPHIE Spectroscopy}
\label{sec:SOPHIE}

We also obtained 113 radial-velocity observations of TOI-2134 with the Spectrographe pour l’Observation des Phénomènes des Intérieurs stellaires et des Exoplanètes (SOPHIE: \citealt{Perruchot2008}) between BJD 2459082 and 2459894 (2020 August 20 to 2022 November 10). SOPHIE is a stabilized \'echelle spectrograph dedicated to high-precision RV measurements in optical wavelengths (387 to 694 nm) on the 193cm Telescope at the Observatoire de Haute-Provence, France \citep{Bouchy2009}. We used the SOPHIE high resolution mode (resolving power $R=75,000$) and the fast mode of the CCD reading. The standard stars observed at the same epochs using the same SOPHIE mode did not show significant instrumental drifts. Depending on the weather conditions, the exposure times for TOI-2134 ranged from 4.5 to 30 minutes (average of 11 minutes) and their SNR per pixel at 550 nm  ranged from 21 to 77 (average of 54). Five exposures showed a SNR below 40 and were removed. The final dataset therefore includes 108 epochs.

The radial-velocity data were extracted with the standard SOPHIE pipeline using CCFs \citep{Bouchy2013} and including the CCD charge transfer inefficiency
correction. The cross-correlations were made using several numerical masks, characteristic of different types of stars. All produced similar results in terms of RV variations. We finally adopted the RVs derived using a K5-type mask, which provided
the least dispersed results.

Following the method described e.g. in \citet{Pollacco2008} and \citet{Hebrard2008}, we estimated and corrected for the sky background contamination (mainly due to the Moon) using the second SOPHIE fibre aperture, which is targeted 2' away from the first one pointing toward the star. We estimated that 14 of the 108 exposures were significantly polluted by sky background, each time implying a correction below 10 \ms.
The final SOPHIE RVs show variations with a dispersion of 8.2 \ms (35 \ms peak to peak), significantly larger than their typical 2 \ms precision. The FWHM, bisector span and contrast of the CCF were also derived for every observation.
The data are plotted in orange in Fig. \ref{fig:all_rv} (for more information on proxy selection see Section \ref{sec:act,rv}).

\section{Stellar Characterisation}
\label{sec:star}

TOI-2134 is a bright, high-proper motion, mid K-dwarf. As the star falls into a parameter space that is not optimal for several of the common stellar characterisation pipelines, we characterised the system with multiple separate and independent methods.
\begin{table*}

\caption{Stellar parameters derived using the different techniques, addressed in order in Sections \ref{sec:star,mann}, \ref{sec:star,annelies} and \ref{sec:star,sam}.
\label{tab:allstars}}

\begin{tabular}{ccccc}\\
\hline
\hline\\

Parameter & SED vs1 & ARES+MOOG & SPC & SED vs2\\
\\
\hline\\
    $F_{\rm bol}$ [erg cm\textsuperscript{2} s\textsuperscript{-1}] & 1.198$\pm$0.048 & & &\\ [2pt]
    $L_{\star}$ [$L_{\odot}$] & 0.192$\pm$0.009 & & &$0.190^{+0.021}_{-0.022}$ \\[2pt]
	$T_{\rm eff}$ [K]  & 4630$\pm$90 & 4620$\pm$80 & 4600$\pm$50 & $4490^{+60}_{-70}$ \\ [2pt]
    Radius $[R_{\odot}]$ & 0.683$\pm$0.027 & $0.714^{+0.017}_{-0.028}$ & & $0.721^{+0.020}_{-0.021}$ \\[2pt]
	log($g$) [cm s\textsuperscript{-1}] &  & 4.8$\pm$0.3 & 4.7$\pm$0.1 & $5.4^{+0.1}_{-0.5}$ \\[2pt]
	${\rm [Fe/H]}$  & & 0.13$\pm$0.04 & 0.09$\pm$0.08 & 0.1\textsuperscript{*} \\[2pt]
	Mass $[M_{\odot}]$ & 0.70$\pm$0.04 & $0.76^{+0.04}_{-0.02}$ && $0.75^{+0.02}_{-0.02}$ \\ [2pt]
    Microturbulence $\xi_{\rm t}$ [km s\textsuperscript{-1}] & &$0.18 \pm 0.12$ &&  \\[2pt]
    Density $[\rho_{\odot}]$ & 2.20$\pm$0.63 & $2.11^{+0.06}_{-0.10}$ && $1.99^{+0.25}_{-0.20}$\\[2pt]
    Age [Gyr] & & $3.8^{+5.5}_{-2.7}$ && 2\textsuperscript{*}\\[2pt]
    Distance [pc] & & 22.646$\pm$0.015 & & $22.657^{+0.006}_{-0.009}$\\

\rule{0pt}{0ex} \\
\hline
\rule{0pt}{0ex} \\

    \textsuperscript{*}Set as constant in the model

\end{tabular}
\end{table*}

\subsection{Spectral Energy Distribution Analysis}
\label{sec:star,mann}
We estimated stellar luminosity $L_{\star}$, effective temperature $T_{\text{eff}}$, and stellar radius $R_{\star}$ by fitting the stellar energy distribution (SED) of \thisstar\ following the method of \citet{Mann2015}, and using templates instead of the observed spectrum, as described in \citet{Mann2016}. To briefly summarise, we compared available photometry ({\it Gaia}, 2MASS, Tycho and WISE) of the host star to a grid of flux-calibrated spectral templates from \citet{Rayner2009} and \citet{Gaidos2014}. We filled gaps in the spectral templates using PHOENIX BT-SETTL models from \citet{Allard2013}, which also provide an estimate of $T_{\text{eff}}$. We computed the bolometric flux, $F_{\rm{bol}}$, by integrating the output absolutely-calibrated spectrum along wavelength. This gave us $L_{\star}$ when combined with the {\it Gaia} DR3 parallax, which in turn gave us $R_{\star}$ when combined with our estimate of $T_{\text{eff}}$ using the Stefan-Boltzmann relation. We did not correct for the offset in the {\it Gaia} DR3 parallax \citep{Lindegren2021}, but this effect is much smaller than the systematic uncertainties intrinsic to the rest of the analysis.

More details on the uncertainties are given in \citet{Mann2015}. To briefly summarise, uncertainties are incorporated as part of a Monte Carlo framework; we generate a grid of fits by sampling over the choice of template (including interpolating between templates), adjustments to the spectral shape (flux calibration uncertainties), as well as reported uncertainties in the parallax, spectra, and photometry. Two irreducible systematic effects were added separately. The first was for $T_{\text{eff}}$ and is based on comparing model-based temperatures to more empirical estimates from long-baseline optical interferometry \citep{Mann2013}. The second was based on calibration of the zero-points and filter profiles \citep{MannvonBraun2015,MaizApellainz2018}. The final values are shown in Table \ref{tab:allstars} under the SED vs1 column.

As part of the analysis, we derived another estimate of $R_{\star}$ based on the scale factor between the models and the absolutely-calibrated spectrum. This scale factor is $\propto R_{\star}^2/D_{\star}^2$, where $D_{\star}$ is the distance to the star . We combined it with the {\it Gaia} parallax to estimate $R_{\star}$. This effectively is the infrared-flux method \citep{Blackwell1977}, and yielded \mbox{$R_{\star}=0.700\pm0.028 R_{\odot}$}, consistent with our Stefan-Boltzmann fit (\mbox{$R_{\star}=0.683\pm0.027 R_{\odot}$}). 

\subsubsection{Stellar mass from $M_{K_S}-M_{\star}$ relation}

We estimated the mass of the host star using the relation between K magnitude and mass, $M_{K_S}$ and $M_{\star}$, from \citet{Mann2019}. This relation was built using orbits of astrometric binaries, making it empirical. Using $K_S$ photometry from the two-micron all-sky survey \citep{Skrutskie2006} and the {\it Gaia} DR3 parallax, we obtained \mbox{$M_{\star}=0.702\pm0.018M_\odot$}. This $M_{\star}$ value placed the host star at the edge of the \citet{Mann2019} relation, where errors may be underestimated due to a lack of Sun-like stars in the sample and the effects of stellar evolution. We, therefore, adopted a more realistic 5\% uncertainty, as shown in Table \ref{tab:allstars}. 

\subsection{ARES+MOOG with isochrone fitting and SPC}
\label{sec:star,annelies}

We also measured stellar atmospheric parameters directly from the HARPS-N spectra. For this purpose the one-dimensional spectra were shifted to the lab-frame with the DRS RVs and then co-added. The resulting spectrum had an SNR of about 600. We employed the ARES+MOOG\footnote{ARESv2: \url{http://www.astro.up.pt/~sousasag/ares/}; \\ MOOG 2017: \url{http://www.as.utexas.edu/~chris/moog.html}} method to measure the effective temperature, surface gravity, microturbulence and iron abundance (used as a proxy for metallicity). We used the method through the FASMA\footnote{FASMA: \url{http://www.iastro.pt/fasma/index.html}} implementation \citep{Andreasen2017}. It relies on calculating the equivalent widths of a set of isolated iron lines (taken from \citealt{Tsantaki2013}) and using them in the radiative transfer code \texttt{MOOG} \citep{Sneden1973} to obtain the atmospheric parameters by imposing excitation and ionisation equilibrium. The stellar atmospheric models were taken from \citet{Kurucz1993}. Some iron lines were discarded as they gave equivalent-width measurements that were unreasonably large ($>200$\,m\AA) or small ($<5$\,m\AA). We also fixed the microturbulence following \citet{Tsantaki2013}. Finally, we inflated the errors for accuracy and corrected the surface gravity following \citet{Mortier2014}. The final values of $T_{\text{eff}}$, surface gravity log($g$), metallicity [Fe/H], and microturbulence $\xi_{\rm t}$ are shown in Table \ref{tab:allstars} under the ARES+MOOG column. 

After obtaining these atmospheric parameters, we used the code \texttt{isochrones} \citep{Morton2015} to derive mass, radius, age, and distance. We ran the code four times, varying the inputs as well as the used stellar models. The common inputs for all four runs were the {\it Gaia} DR3 parallax, and the photometric magnitudes in bands B, V, J, H, and K. For two runs, we also included the effective temperature and metallicity as measured from the HARPS-N spectra. We chose not to use the spectroscopic surface gravity given its known accuracy issues \citep[see e.g. ][]{Mortier2014}. We used two stellar models (each in two runs): the Dartmouth Stellar Evolution Database \citep{Dotter2008} and the MESA Isochrones and Stellar Tracks \citep[MIST: ][]{Dotter2016}. For our final results, we combined the posterior distributions of all four runs. To combine the posteriors we added them together and corrected for the sample size (as in \citealt{Borsato2019}). We extracted the median and 16th and 84th percentiles as the final value and its errors, as reported in Table \ref{tab:allstars}.

\subsubsection{SPC Pipeline}
We also derived stellar parameters using the Stellar Parameter Classification pipeline \citep[SPC:][]{Buchhave2012, Buchhave2014}. The high signal to noise ratio needed to extract precise RVs means that these spectra are more than adequate for deriving stellar parameters. We ran the SPC analysis on each individual spectrum and calculated the weighted average of the individual spectra.
The weights are computed from the normalised CCF peak heights from the observed spectrum and the best matched template (model) spectrum. Higher CCF peaks indicate a better match between the model and the observations. The normalisation leads to a CCF peak height of 1 for autocorrelation. While the SNR of the observed spectra could also be used as the weighting factor, the CCF peak height better incorporates the relationship between data and model.
The results are show in Table \ref{tab:allstars} under the SPC column. We also computed $v{\rm sin}(i) < 2~{\rm km~s^{-1}}$.
The formal uncertainties take into account the model uncertainties, which primarily stem from systematics in the ATLAS Kurucz stellar models and degeneracies between the derived parameters when trying to compare observed spectra to model spectra \citep[see][]{Buchhave2012, Buchhave2014}. The parameters from SPC agree well with the results from ARES+MOOG within the uncertainties. 

\subsection{Spectral Energy Distribution Analysis with Isochrone Fitting}
\label{sec:star,sam}
We have also computed an estimate of $R_\star$ and $T_{\rm eff}$ using the SED fitting method presented in \citet{Morrell2019a, Morrell2020a}.
This method compares multiband photometry placed across the stellar SED with synthetic photometry, generated from the \mbox{BT-Settl CIFIST} \citep{Allard2012a} atmosphere grid, and diluted using the distances of \citet{BailerJones2021a}.
By best matching the area beneath the SED and the overall shape of the SED, we determined the luminosity $L_{\rm SED}$ and temperature $T_{\rm SED}$ respectively -- which together unambiguously define $R_\star$. 
Unlike the method presented in Section \ref{sec:star,mann}, which makes use of spectroscopic templates for the measurement of $T_{\rm eff}$, this method self-consistently measures both $T_{\rm eff}$ and $R_\star$ using only photometry and distances, effectively providing an alternate measure of temperature to the other methods.

\begin{table}
\centering

\caption{Stellar parameters of TOI-2134.
\label{bigtable}\label{tab:star}}

\begin{tabular}{c@{\hskip6pt}c@{\hskip7pt}c}

\rule{0pt}{0ex} \\
\hline
\hline
\rule{0pt}{0ex} \\
Parameter & Value & Source \\
\rule{0pt}{0ex} \\
\hline
\rule{0pt}{0ex} \\
    Name           & TOI-2134   &   TESS Project\textsuperscript{*}\\
                    & TIC 75878355 & \cite{Stassun2019} \\
                   & G204-45    &   \cite{Giclas1979}\\
	RA [h:m:s]      & 18:07:44.52 &  \cite{G32020} \\
	DEC [d:m:s]     & +39:04:22.54 & \cite{G32020}  \\
	Spectral type   & K5V  & \cite{Stephenson1986} \\
	$m_V$ [mag]   & 8.933$\pm$0.003 & TESS Project\textsuperscript{*} \\
    $m_J$ [mag]   & 6.776$\pm$0.023 & TESS Project\textsuperscript{*} \\
    $m_K$ [mag]   & 6.091$\pm$0.017 & TESS Project\textsuperscript{*} \\
    $(B-V)$ [mag]   & 1.192$\pm$0.033 & TESS Project\textsuperscript{*} \\
	Parallax [mas]  & 44.1087$\pm$  0.0144 & \cite{G32020} \\
	Distance [pc]   & 22.655$\pm$0.007 & this work\\
    Proper motion [mas/yr] & 288.257$\pm$0.016 & \cite{G32020}\\
    $L_{\star}$ [$L_{\odot}$] & 0.192$\pm$0.008 & this work\\
    $F_{bol}$ [erg cm$^2$ s$^{-1}$] & 1.198$\pm$0.048 & this work\\
	$T_{\rm eff}$ [K]  & 4580$\pm$50  & this work \\
	log($g$) [cm s$^{-1}$]& 4.8$\pm$0.3 & this work \\
	${\rm [Fe/H]}$  & 0.12$\pm$0.02 & this work \\
	Mass $[M_{\odot}]$ & 0.744$\pm$0.027 & this work \\
	Radius $[R_{\odot}]$ & 0.709$\pm$0.017 & this work \\
	Density [$\rho_{\odot}$] & 2.09$\pm$0.10 & this work \\
    Age [Gyr] & $3.8^{+5.5}_{-2.7}$ & this work \\
	$v$sin$(i)$ [km~s$^{-1}$]      & 0.78$\pm$0.09 & this work \\
	$<\log{R'_{\rm HK}}>$ & -4.83$\pm$0.45 & this work \\ [2pt]
	P$_{\rm rot}$ [days] & \Pk$^{\Pku}_{\Pkd}$ & this work \\

\rule{0pt}{0ex} \\
\hline
\rule{0pt}{0ex} \\

\end{tabular}

\textsuperscript{*}See ExoFOP: \url{https://exofop.ipac.caltech.edu/tess/target.php?id=75878355}

\end{table}

For this fitting we used the $G_{\rm BP}$ and $G_{\rm RP}$ bands from {\it Gaia} DR3 \citep{Gaia2016, Gaia2023}, the J, H, and K bands from 2MASS \citep{Skrutskie2006}, and the W1, W2, and W3 bands from AllWISE \citep{Wright2010}. 
As with \citet{Morrell2019a}, we adopted a floor value of 0.01 mag, corresponding to about 1\%, for the photometric uncertainty for all bands. 
The parameters resulting from our fitting are shown in the SED vs2 column in Table \ref{tab:allstars}.
At first glance, $T_{\rm eff}$ and $R_\star$ from this method are inconsistent with the other determinations. However, the resulting $L_\star$ from these parameters is consistent with that described in Section \ref{sec:star,mann}, supporting the validity of both sets of parameters. We considered the possibility of extinction contributing to the aforementioned difference, however the star is close enough that extinction should be negligible. Moreover the extinction required to match the results of Section \ref{sec:star,mann} is 0.1, which is too large to be probable.
Furthermore, the measurement of $R_\star$ using this method is consistent with the secondary, infrared flux-based method determination from Section \ref{sec:star,mann}. 
From our study, the $G_{\rm BP}$ and $G_{\rm RP}$ bands appear to be sampling a redder SED than the bands at longer wavelengths, resulting in a cooler measured $T_{\rm SED}$. 
Given that the photometric data were not contemporaneous, with the visible and IR photometry being 5-10 years separated, it is possible for the observed SED to have changed over this intervening period. 
Though, as we can find no quality issues or physical reason for this discrepancy, the fitting for our parameter determinations for this section did employ the $G_{\rm BP}$ and $G_{\rm RP}$ bands.

We then determined the stellar mass $M_\star$ using the PARSEC 1.2S isochrones \citep{Marigo2017,Bressan2012,Chen2014,Chen2015,Tang2014,Pastorelli2019}. 
We used \mbox{\texttt{CMD 3.7}}\footnote{CMD 3.7: \url{http://stev.oapd.inaf.it/cgi-bin/cmd}} to generate evolution tracks at a metallicity of \mbox{[M/H] = 0.1}, which is in line with the value determined in Section \ref{sec:star,annelies}. 
Given that the ARES+MOOG age estimation places the star on the main sequence, we interpolated the 2 Gyr isochrone to estimate the $M_\star$ at our measured $L_\star$ and its uncertainty bounds, also shown in Table \ref{tab:allstars}.
We note that, due to not having access to the posterior for distance and instead just assuming it to be Gaussian, the uncertainty bounds for luminosity, mass, and stellar density from this method are likely to be overestimated.\\

Overall, all analysis agree with each other within their uncertainties. For the scope of this work, we characterised TOI-2134 via the mean of all the computed values weighted by the inverse of their errors, as compiled in Table \ref{tab:star}. Their uncertainties are computed as the standard deviation between measurements in each method, to avoid improper averaging down of systematic effects.

\begin{figure*}
	\includegraphics[width=17cm]{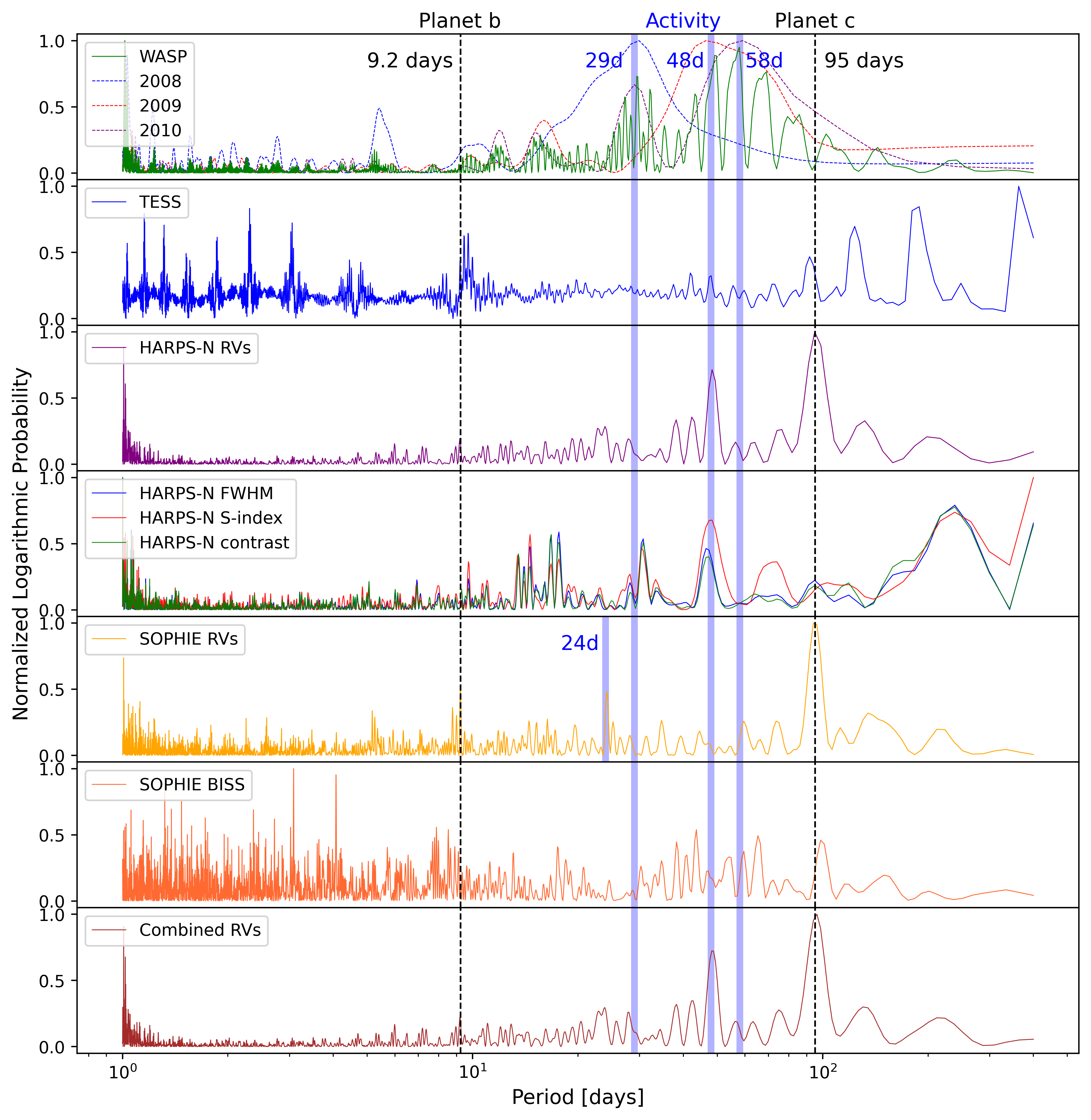}
    \caption{Set of BGLS periodograms of the acquired data plotted as period vs logarithmic probability normalised to 1. From the top, WASP photometry in the solid green with yearly seasons in blue for 2008, red for 2009 and purple for 2010 as dashed lines, TESS photometry, HARPS-N RVs, HARPS-N activity proxies (FWHM, S-index and contrast in respectively blue, red and green), SOPHIE RVs, SOPHIE activity proxy (BISS), and the combined SOPHIE and HARPS-N RVs. The dashed black lines represent the periods of the two planets at 9.2 and 95 days. The blue bands indicate the possible stellar rotational signals at 29, 48 and 58 days.}
    \label{fig:full_per}
\end{figure*}

\begin{figure*}
	\includegraphics[width=17cm]{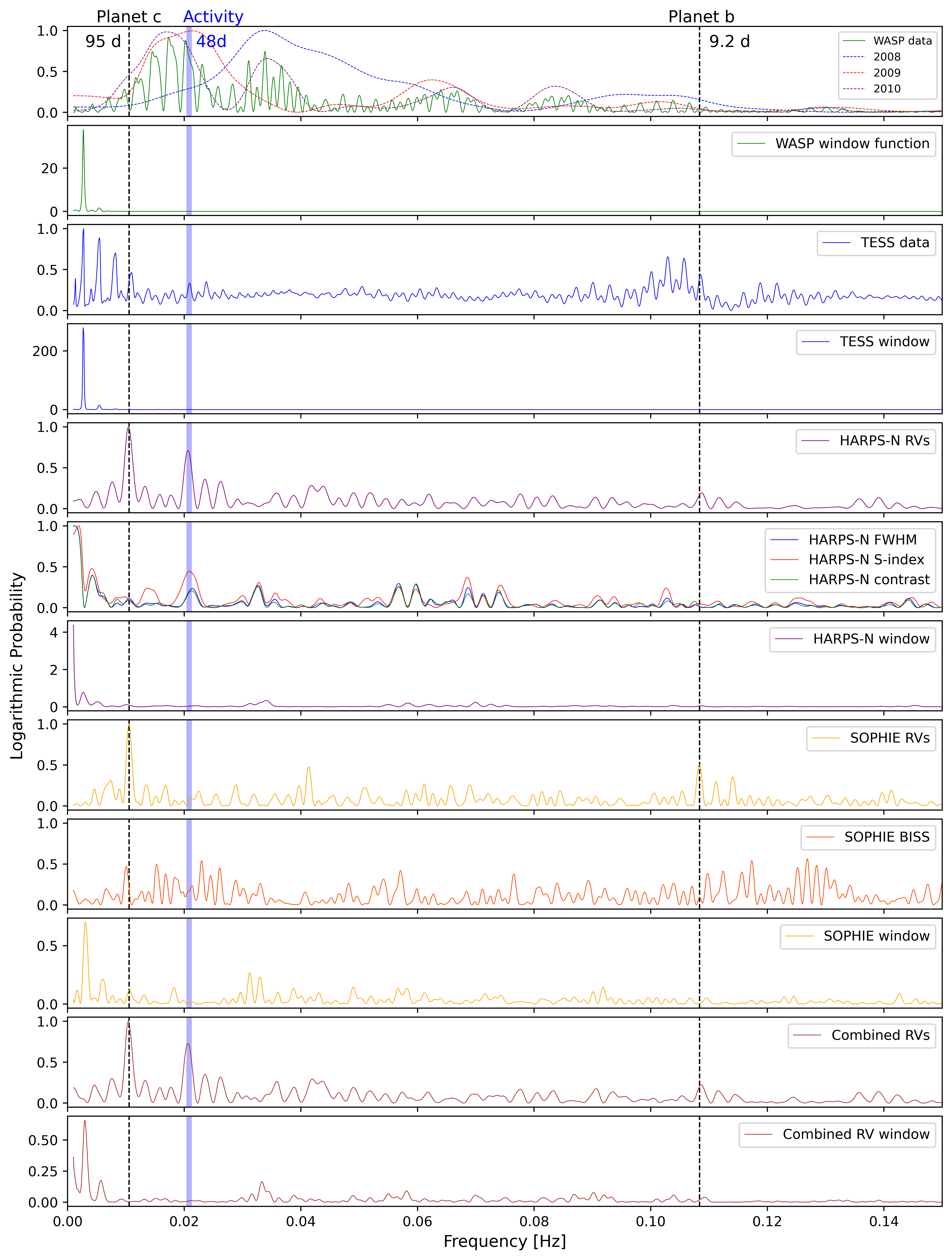}
    \caption{Same set of BGLS periodograms as Fig \ref{fig:full_per} in frequency space. The window functions for each dataset are also included. The dashed black lines indicate the periods of the two planet candidates. The blue band shows the stellar rotational period.}
    \label{fig:full_window}
\end{figure*}
\section{Stellar Activity Signal}
\label{sec:activity}
We conducted a thorough preliminary analysis of the available data in order to search for and to best characterise the stellar activity-induced signals in both the photometric and the spectroscopic observations.

To begin with, the projected rotational velocity $v$sin$(i)$ of TOI-2134 was determined to be  <2 \kms from the HARPS-N spectra (as mentioned in the previous Section), and 1.5$\pm$1.0 \kms from the SOPHIE cross-correlation functions (following the method in \citealt{Boisse2010}). No more precise measurement could be derived from the spectra.
We therefore calculated a minimum stellar rotation period $P_{\rm rot, min}$ associated to the lower maximum limit of $v$sin$(i)$ as:
\begin{equation}
    P_{\rm rot, min} = \frac{2\pi R_{\star}}{v {\rm sin}(i)} \approx 23 \:\mathrm{days}.
    \label{eq:Prot}
\end{equation}

Using the method described in \cite{Noyes1984}, we computed the average $\log{R'_{\rm HK}}$ to be -4.83$\pm$0.45 from the S-index measurements taken by HARPS-N. There was significant scatter in the S-index measurements which degraded the quality of the results, but the empirical relations of \cite{Noyes1984} yielded a stellar rotation period of $\sim$42 days.\\

To better identify the stellar rotational period we performed a periodogram analysis.

\subsection{Photometry}
\label{sec:act,photo}
We computed the Bayesian Generalised Lomb-Scargle (BGLS) periodograms \citep{Mortier2015} for both the WASP and the TESS photometric data, shown respectively in green and blue in the first and second rows of Fig. \ref{fig:full_per}. The same periodograms in frequency space, alongside their window functions are shown in Fig. \ref{fig:full_window}.
The TESS data showed a forest of peaks at $\sim$9.2 days (highlighted by a black dashed line), which is generated by the repeated transits of the inner planet. As expected given the detection of no transits due to lower precision, the WASP periodogram had no power around this period.
It instead showed two significant forests of peaks centred around $\sim$29 and $\sim$58 days (shown as blue bands in Fig. \ref{fig:full_per}), which were originally attributed to the stellar rotational period, but could also be generated by the moon cycle. To further investigate this, we also plotted the BGLS periodograms of each yearly season of WASP, as shown in the first row of Fig. \ref{fig:full_per} as blue, red and purple dashed lines. The BGLS periodograms of the two later years also presented a significant peak at 58 days, but the 2008 data did not. Instead, its most significant peak was at 29 days. A peak at $\sim$29 days was also present in 2010, but not in 2009. While some of the discrepancies could be attributed to differing coverage, these result hinted at either a different lunar contribution over the different seasons, or at evolving surface inhomogeneities structure trends over the years, possibly related to a stellar magnetic cycle. After alias analysis, we found that the 29 days forest of peaks in the full periodogram can be explained as the extended aliases generated by the 365 days period.
The WASP data span over $\sim$850 days. SOPHIE radial velocities (taken 10 years later) also cover a similar stretch of time. Therefore, assuming these signals are stellar, we can expect the structure of surface inhomogeneities that allow us to detect stellar rotational period in periodogoram analyses to also evolve during the three years of radial-velocity data. This evolution could be the reason behind the difficulties constraining the stellar rotational period in the further RV analyses.

\subsection{Radial-Velocity Data and Proxies}
\label{sec:act,rv}
We conducted a full periodogram analysis of the spectroscopic data. The last five rows of Fig. \ref{fig:full_per} show the BGLS periodograms of, in order, the HARPS-N RVs, the HARPS-N derived proxies (FWHM, S-index and contrast), the SOPHIE RVs, the SOPHIE derived activity indicator (bisector span, or BISS), and the combined RV data. We were able to combine the RVs with a simple offset, as they are derived from similar wavelength windows and therefore are probing the same section of the stellar spectra.  The same periodograms in frequency space, alongside their window functions, are once again shown in Fig. \ref{fig:full_window}.

Although the star showed significant variation in the activity indicators, and the average $\log{R'_{\rm HK}}$ also classified the star as moderately active, both sets of RVs had little to no correlation to their activity indicators. The specific reason for this lack of correlation is ultimately beyond the scope of this paper, as the activity indicators were only used as a starting point to the analysis, but we propose some possible origins. As a first most likely option, the Keplerian signals introduced by the planets in the system are large enough to "muddle" the correlation to activity indicators. In this case, the RV amplitude of the stellar activity computed in the next Sections is shown to be comparable to the amplitude of the RV signals generated by the planets. It is likely that these signals are significant enough to prevent a clean correlation between RVs and activity indicators (which only map the variations induced by stellar activity).
To test this, we also computed the correlation between the activity indicators and the RVs after subtracting the best-fit Keplerian models computed in Section \ref{sec:results}. While the correlation did improve by a factor of 2, they still remained low. So other reasons may be considered. As an example, the stellar rotation axis inclination angle with respect to the observer can influence the strength of this correlation, weakening it for unfavourable line-of-sights: as the the stellar rotational axis becomes parallel to the observer line-of-sight, the signal from active regions coming in and out of view becomes less rotationally modulated. At the same time, in late K-dwarfs convective redshift may in some cases prevail against blueshift. This can happen either due to an opacity effect (like in M-type stars), or if most of the photospheric absorption lines used for RV measurements form in regions of convective overshoot \citep{Norris2017}. \cite{Costes2021} notes that a possible explanation for low correlation between radial velocities and activity proxies, as is the case for our target, is that the convective blue- and redshifts are "cancelling" one another. The possibility of a temporal lag \citep{CollierCameron2019} between the radial velocities and the proxies was also considered, but a visual inspection of their timeseries did not strongly support this possibility.

For our analysis we nevertheless selected and plotted the indicators with the strongest correlation to their RVs. For HARPS-N we selected the S-index, the FWHM and the contrast. Their Spearman's rank correlation coefficients with the RVs were computed to be 0.15, 0.11 and -0.12 respectively. For SOPHIE we selected only the bisector span, with correlation coefficient of -0.16, as the FWHM and contrast seem to be affected by instrumental systematics.\\

\begin{figure}
    \includegraphics[width=\columnwidth]{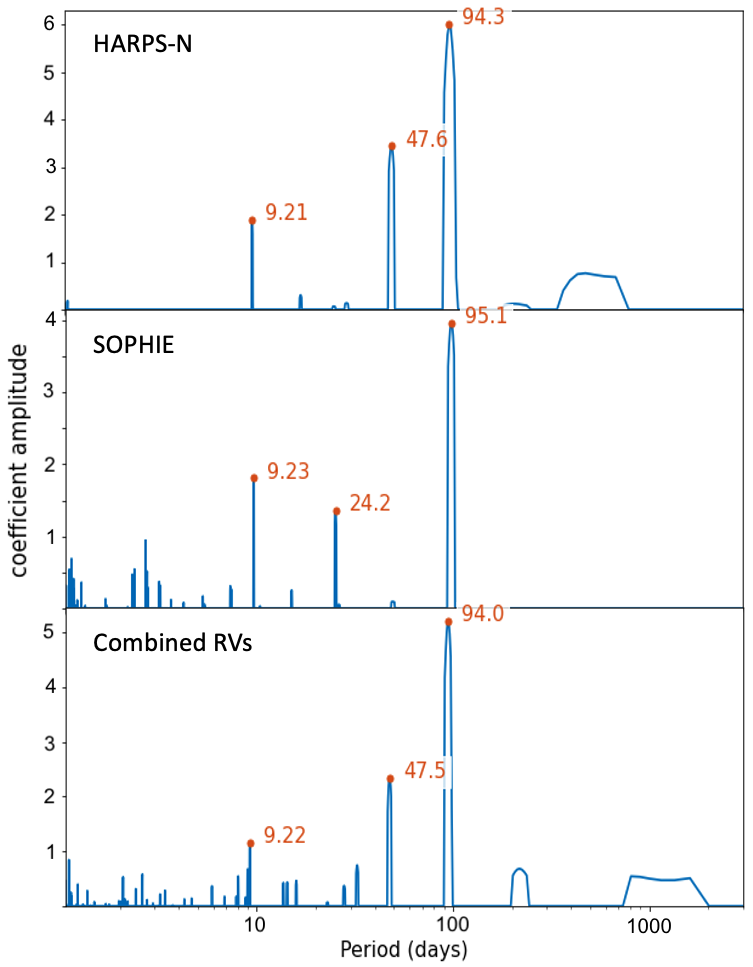}
    \caption{$\ell$1 periodograms of from top to bottom HARPS-N, SOPHIE and combined RVs. The periods of the major identified signals are highlighted in red.}
    \label{fig:l1}
\end{figure}
\begin{figure}
    \centering
    \includegraphics[width=7.5cm]{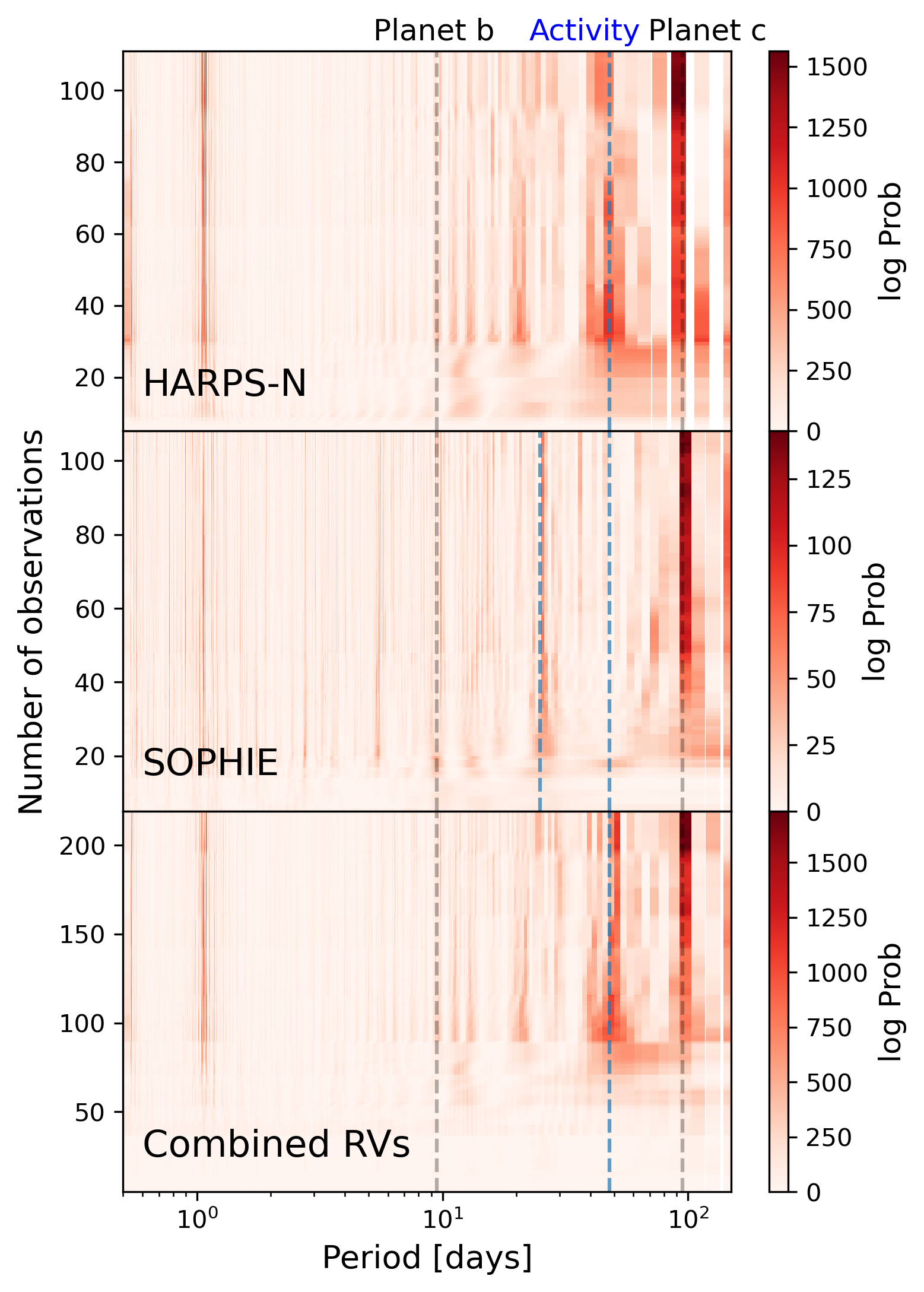}
    \caption{From the top, stacked BGLS periodograms of HARPS-N, SOPHIE and combined RVs. The blue dashed blue lines identify 48 days (and 25 days for SOPHIE data). The dashed grey lines show 9.23 and 95 days, the proposed periods of the two planets.}
    \label{fig:sbgls}
\end{figure}

While the BGLS periodograms of the radial-velocity datasets did not show clear peaks for the inner planet, there was a strong periodic signal at $\sim$95 days (shown as a black dashed line) shared between HARPS-N and SOPHIE RVs that was not present in any of the HARPS-N stellar activity proxies. The SOPHIE bisector does have a peak at $\sim$100 days, but its normalised logarithmic probability is comparable to most other peaks in the periodogram and therefore does not have a strong relevance.
This preliminary analysis suggested a period of $\sim$95 days for the mono-transiting planet detected by TESS. This signal presented minorly relevant yearly aliases at 129 days and 75 days in both the HARPS-N only and the combined data, which could be easily discarded in the analysis of the periodogram. No statistically significant yearly aliases arise for the 95 days signal in the SOPHIE data.
The only major peak of both HARPS-N radial velocities and of all its activity indicators was centred around 48 days (shown as a blue band). In the HARPS-N data, we could also see some of the yearly aliases of this signal, at 42 and 38 days. This peaks were only  moderately relevant and could be easily identified.
No such signal can be found in either SOPHIE RVs or its indicator. On the other hand, SOPHIE data presented a minor peak at $\sim$24 days, half of the HARPS-N value. This disagreement could be due to the different sampling and observing strategies between the two observatories. Further alias analysis showed that 24 days was also a yearly alias of 48 days.
The 48 days period, although not in perfect agreement, is compatible with the longer modulation in the WASP data, especially given the fact that the data in each season only span just more than twice this period.\\

To further analyse the signals within the spectroscopic datasets, we have also included an $\ell$1 periodogram\footnote{Available at \url{https://github. com/nathanchara/l1periodogram}}  analysis with correlated noise \citep{Hara2017, Hara2021py}, as shown in Fig. \ref{fig:l1}. This periodogram formulation was first devised to overcome the distortions in the residuals that arise when fitting planets one by one, and can help isolate the most relevant signals in a dataset.
Once again, HARPS-N and SOPHIE radial velocities on their own, as well as their combination, all showed a clear peak at $\sim$95 days. Similarly, the $\ell$1 periodograms of HARPS-N and SOPHIE both also peaked at $\sim$9.2 days. The $\ell$1 periodogram is also able to isolate the signal of the inner planet in the combined RV dataset. Regarding the possible stellar rotation period, HARPS-N data again showed a clear modulation at $\sim$48 days, while the strongest peak in SOPHIE not attributed to planetary signals was at half that value. The $\ell$1 periodograms have therefore re-confirmed the previous results from the BGLS analysis and have allowed for a clearer understanding of the SOPHIE data.\\

\begin{figure}
	\includegraphics[width=\columnwidth]{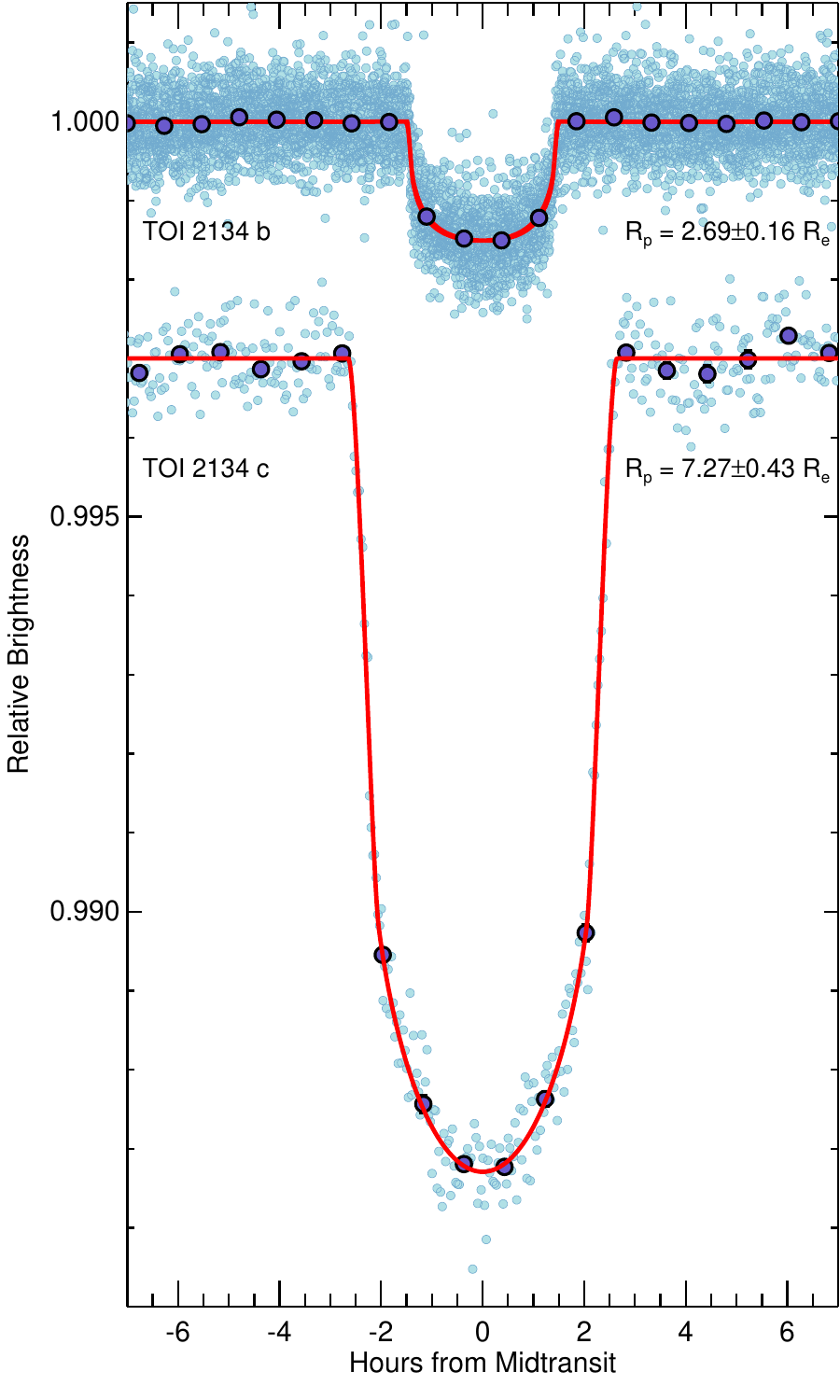}
    \caption{Phase-folded TESS light curves of TOI-2134b and c. Faint blue points are individual TESS two-minute cadence measurements, bold darker blue points are data binned in orbital phase, and the red curves are the best-fit transit models. The error bars on the binned points are smaller than the symbols. For the transit of TOI-2134c, we artificially offset the out-of-transit flux measurements for improved visibility. }
    \label{fig:TESS_fits}
\end{figure}
Finally, to probe the coherence of these signals, we plotted the Stacked Bayesian Generalised Lomb-Scargle periodograms \citep{Mortier2015, Mortier2017} of the three sets of RV data in Fig. \ref{fig:sbgls}. The Stacked BGLS periodogram was developed to better identify the signals that are generated by stellar activity. Planetary signals are coherent in nature, meaning their probability should consistently increase with increasing number of observations. Signals produced by stellar activity are incoherent, meaning that their probability will change and oscillate. Fig. \ref{fig:sbgls} clearly showed that the signals indicated by the blue vertical lines (respectively 24 and 48 days, as identified by the $\ell$1 periodograms) were incoherent. They therefore could not be attributed to planets and were more likely generated by stellar activity. The 9.2 and 95 days signals, highlighted by the grey dashed lines, showed more coherent trends. With the exception of a 1 day alias, no other major signals could be identified.

\section{Transit Photometry}
\label{sec:transit}
We then performed an analysis to determine the best-fit transit parameters and uncertainties for the two planet candidates orbiting TOI-2134. We modelled the TESS photometry (after systematics correction and flattening as described in Section \ref{sec:TESS}) with \citet{MandelAgol2002} transit models. Our model included three parameters describing the host star (its mean density, and both linear and quadratic $q_1$ and $q_2$ limb-darkening coefficient parametrisations sampled following \citealt{Kipping2013}). The inner planet TOI-2134b was described by six parameters (its orbital period, time of transit, orbital inclination, the logarithm of the planet/star radius ratio $\log{R_p/R_\star}$, and combinations of the eccentricity and argument of periastron of the planet $\sqrt{e}\cos{\omega_{\rm p}}$ and $\sqrt{e}\sin{\omega_{\rm p}}$, which will be further explained in Section \ref{sec:rv}). The transit of the outer planet TOI-2134c was described by four parameters (time of transit, transit duration, impact parameter, and the logarithm of the planet/star radius ratio). Finally, we included two parameters characterising the dataset itself (a constant flux offset and the white noise level).

\begin{table}
\centering
\caption{Results and uncertainties of the planetary parameters for the photometry analysis described in Section \ref{sec:transit}. \label{tab:phot}}
\begin{tabular}{ccc}
\hline
\hline
\rule{0pt}{0ex} \vspace{-0.2cm} \\
Parameter & Value\\
\rule{0pt}{0ex} \vspace{-0.2cm} \\
\hline
\rule{0pt}{0ex} \vspace{-0.2cm} \\
    Radius ratio $(R_{\rm b}/R_\star)$ & \rprstb$\pm$ \urprstb \\
    Orbital period $P_{\rm b}$~[days] & \perplb$\pm$\uperplb\\
    Time of transit $t_{\rm 0,b}$~[BJD] & \ttransitb$\pm$\uttransitb\\ 
    Orbital inclination $i_{\rm b}$~[deg] & \inclb$\pm$ \uinclb \\ 
    Transit impact parameter $b_{\rm b}$ & \impb$\pm$\uimpb  \\
    Radius ratio $(R_{\rm c}/R_\star)$ & \rprstc$\pm$ \urprstc\\
    Time of transit $t_{\rm 0,c}$~[BJD] & \ttransitc$\pm$\uttransitc\\ 
    Transit impact parameter $b_{\rm c}$ & \impc$\pm$ \uimpc \\
\rule{0pt}{0ex} \vspace{-0.2cm} \\
\hline
\end{tabular}	
\end{table}
\subsection{Selection of Priors}
We imposed an informative Gaussian prior on the stellar density based on our analysis of the stellar parameters. 
All other parameters were bound by uniform priors. We restricted the inclination of planet b to be less than 90$^\circ$ and the impact parameter of planet c to be greater than 0 (to avoid the degeneracy for transit configurations with inclinations greater 90$^\circ$). 
We restricted $\sqrt{e}\cos{\omega_{\rm p}}$ and $\sqrt{e}\sin{\omega_{\rm p}}$ to be in the interval [-1,1] (as necessary as per their definition), and the impact parameters (in the case of TOI-2134b after conversion from inclination) to be in the range [0,1+$R_p/R_\star$] (requiring the planets transit the star). $\log{R_p/R_\star}$ was allowed to vary in the range $[-\infty, 0]$ (planets must be smaller than the host star), and $q_1$ and $q2$ in the range [0,1] following \citet{Kipping2013}. All other parameters with uniform priors were allowed to explore the range  $[-\infty, \infty]$\footnote{To be precise python defines its minimum and maximum float values to specific numbers, so these are actually uniform priors between \mbox{[-1.7976931348623157$\cdot10^{308}$, 1.7976931348623157$\cdot10^{308}$]}}.

\subsection{Transit Results}
We explored the parameter space using a Markov Chain Monte Carlo (MCMC) algorithm with a Differential Evolution sampler \citep{TerBraak2006}. We simultaneously evolved 100 chains for 100,000 steps each, discarding the first 30,000 as burn-in. We assessed convergence by calculating the Gelman-Rubin statistic and found values less than 1.006 for all parameters. Our best-fit models are phase-folded and plotted in Figure \ref{fig:TESS_fits} and the results of our planetary fit are given in Table \ref{tab:phot}. We chose to initially not derive eccentricity, angle of periastron and period for the outer planet candidate, given the mono-transit. Those parameters will be extracted in a second step we discuss in Section \ref{sec:bimecc}.
The multiple transits of the inner planet allow us to precisely measure its period and planet-to-star radius ratio. The radius ratio of TOI-2134c is also constrained to over 100$\sigma$.

\begin{table*}
\centering
\caption{Results from the three Gaussian Process regression analysis. We include the priors applied to each parameter. In order the HARPS-N RVs only, and the SOPHIE RVS only results, followed by the combined HARPS-N and SOPHIE data results (used for all further analysis). We abbreviated uniform priors as $\mathcal{U}$, Gaussian priors as $\mathcal{G}$ and Jeffreys' priors as $\mathcal{J}$. We only show the results obtained for the high eccentricity case, as addressed in Section \ref{sec:bimecc}. \label{tab:separategp}}

\begin{tabular}{ccccc}

\hline
\hline\\[-5pt]

Parameter & Prior & HARPS-N RVs & SOPHIE RVs & Combined RVs \\ [4pt]

\hline\\[-5pt]

    GP Amplitude $\theta_1$ [\ms] & $\mathcal{U}[0,20]$ & 4.24$_{-0.59}^{+0.81}$ & 5.52$_{-0.68}^{+0.98}$  & \Kk$_{\Kkd}^{\Kku}$\\ [2pt]
    GP Timescale $\theta_2$ [days] & $\mathcal{J}[0,100]$ & 31.84$_{-10.36}^{+9.93}$ & 10.15$_{-7.99}^{+22.51}$ & \Evok$_{\Evokd}^{\Evoku}$\\ [2pt]
    GP Period $\theta_3$ [days] & $\mathcal{G}[48,10$] & 45.85$_{-4.84}^{+4.89}$ & 38.89$_{-14.17}^{+13.99}$ & \Pk$_{\Pkd}^{\Pku}$ \\ [2pt]
    GP Smoothness $\theta_4$ & $\mathcal{G}[0.5,0.05]$ & 0.48$_{-0.05}^{+0.05}$ & 0.48$_{-0.05}^{+0.06}$ & \Harmk$_{\Harmkd}^{\Harmku}$\\ [2pt]
    Jitter [\ms] & $\mathcal{U}[0,2]$ & 0.69$_{-0.12}^{+0.13}$ & 0.82$_{22}^{+0.24}$ & \jit$_{\jitd}^{\jitu}$\\ [2pt]
    SOPHIE - HARPS-N Offset [\ms] & $\mathcal{U}[-5,5]$ & & & \offset$_{\offsetd}^{\offsetu}$ \\ [2pt]
    Orbital period $P_b$~[days] & $\mathcal{G}[9.2292004,0.0000063]$ & 9.22923$_{-0.00003}^{+0.00004}$ & 9.2292$_{-0.0001}^{+0.0002}$ & \Pb$_{\Pbd}^{\Pbu}$\\ [2pt]
    RV Amplitude $K_b$ [\ms] & $\mathcal{U}[0,20]$ & 3.01$_{-0.32}^{+0.32}$ & 4.13$_{-0.87}^{+0.84}$ & \Kb$_{\Kbd}^{\Kbu}$\\ [2pt]
    $S_{\rm k,b}$ & $\mathcal{U}[-1,1]$ & -0.04$_{-0.08}^{+0.06}$ & 0.21$_{-0.09}^{+0.08}$ & \Skb$_{\Skbd}^{\Skbu}$\\ [2pt]
    $C_{\rm k,b}$ & $\mathcal{U}[-1,1]$ & 0.22$_{-0.09}^{+0.06}$ & 0.21$_{-0.09}^{+0.07}$ & \Ckb$_{\Ckbd}^{\Ckbu}$\\ [2pt]
    Time of periastron $t_{\rm peri,b}$ [BJD] & $\mathcal{G}[2459408.22, 0.50]$ & 2459407.71$_{-0.38}^{+0.46}$ & 2459407.55$_{-0.44}^{+0.43}$ & \tperib$_{\tpericd}^{\tpericu}$\\ [2pt]
    Orbital period $P_c$~[days] & $\mathcal{U}[75,150]$ & 94.71$_{-1.11}^{+1.17}$ & 94.86$_{-0.83}^{+1.13}$ & \Pc$_{\Pcd}^{\Pcu}$\\ [2pt]
    RV Amplitude $K_c$ [\ms] & $\mathcal{U}[0,20]$ & 11.92$_{-1.82}^{+1.82}$ & 10.28$_{-2.94}^{+2.99}$ & \Kc$_{\Kcd}^{\Kcu}$\\ [2pt]
    $S_{\rm k,c}$ & $\mathcal{U}[-1,1]$ & -0.65$_{-0.07}^{+0.11}$ & 0.69$_{-0.10}^{+0.29}$ & \Skc$_{\Skcd}^{\Skcu}$\\[2pt]
    $C_{\rm k,c}$ & $\mathcal{U}[-1,1]$ & 0.42$_{-0.21}^{+0.13}$ & 0.41$_{-0.32}^{+0.66}$ & \Ckc$_{\Ckcd}^{\Ckcu}$\\ [2pt]
    Time of periastron $t_{\rm peri,c}$ [BJD] & $\mathcal{U}[2459678.5,2459773.5]$ & 2459724.33$_{-2.53}^{+3.27}$ & 2459731.05$_{-8.86}^{+4.07}$ & \tperic$_{\tpericd}^{\tpericu}$\\[4pt]
    
\hline

\end{tabular}	
\end{table*}
\section{Radial-Velocity Analysis}
\label{sec:rv}
To analyse the radial velocities we used the new code \texttt{MAGPy\_RV}\footnote{Available at \url{https://github.com/frescigno/magpy_rv}}. \texttt{MAGPy\_RV} is a pipeline for Gaussian Process regression with an affine invariant MCMC parameter space searching algorithm (as defined in \citealt{Foreman-Mackey2013}).

Gaussian Processes have been extensively employed in astrophysical literature to successfully model stellar activity-induced variations and instrumental noise in both radial-velocity and photometric measurements (e.g., \citealt{Haywood2014, Rajpaul2015, Faria2016, Serrano2018, Barros2020}).\\

We modelled the RV data as a combination of two planetary signals in the form of Keplerians (for the two transiting objects), and the stellar activity in the form of a Quasi-Periodic kernel. We selected the Quasi-Periodic Kernel defined in \citet{Haywood2014} with the inclusion of a white noise "jitter" term, in the form
\begin{equation}
    k(t_n,t_m) = \theta_1^2 \cdot \exp\left[ -\frac{|t_n - t_m|^2}{\theta_2^2} - \frac{\sin^2 \left(\frac{\pi \cdot |t_n - t_m|}{\theta_3} \right)}{\theta_4^2} \right] + \delta_{n,m}\beta^2,
    \label{eq:QP}
\end{equation}
in which $t_{n}$ and $t_{m}$ are two datapoints, the four hyperparameters $\theta$s are in order the maximum amplitude, the timescale over which the quasi-periodicity evolves, the period of the periodic variation (mapping the stellar rotation), and the "smoothness" of the fit (its amount of high-frequency structure) also often referred to as the harmonic complexity. The "jitter" term is represented by the delta function and $\beta$ can be thought of as the contribution to the RVs from the precision on the spectrograph.

While eccentricity $e$, and planetary angle of periastron $\omega_{\rm p}$ were used within the Keplerian model, when iterating in the MCMC algorithm we instead took steps in a different set of variables $S_{k}$ and $C_{k}$, defined as
\begin{equation}
\begin{gathered}
    S_k = \sqrt{e} \sin{\omega_{\rm p}},\\
    C_k = \sqrt{e} \cos{\omega_{\rm p}}.
\end{gathered}
    \label{eq:SkCk}
\end{equation}
As explained in \citet{Eastman2013}, this reparameterisation avoids a boundary condition at zero eccentricity, allowing for a better sampling around zero while maintaining the overall prior flat over eccentricity.

The Keplerian models also depended on time of periastron passage $t_{\rm p}$, rather than the time of transit $t_{\rm 0}$, derived by transit photometry. However, the two variables are linked via the following equation
\begin{equation}
    t_{\rm p} = t_{\rm 0} -\frac{P}{2\pi} \cdot \left[ E_{\rm tr} -e \cdot \sin\left(E_{\rm tr}\right)\right],
    \label{eq:tr_to_peri}
\end{equation}
in which $P$ is the orbital period of the considered planet, $e$ its eccentricity and the eccentric anomaly $E_{\rm tr}$ is computed from the argument of periastron and the eccentricity as
\begin{equation}
    E_{\rm tr} = 2\arctan\left[\sqrt{\frac{1-e}{1+e}} \cdot \tan\left(\frac{\pi - 2\omega_{\rm p}}{4}\right)\right].
    \label{eq:Etr}
\end{equation}
\\

We conducted our investigation on the combined HARPS-N and SOPHIE dataset, as well as on the two datasets separately. Once again we were able to combine the two RV datasets with a simple offset parameter and could use a single GP to describe both because they have comparable jitters and they are derived by similar spectral windows in the optical range. Therefore, they are expected to map the same physical processes and to be sensitive to Doppler-shift in the same way.

\subsection{Selections of Priors}
\label{sec:gpprior}
In this section we describe the choices of priors for the analysis of the RV data. The same priors are used for all three analyses. They are also summarised in Table \ref{tab:separategp}.

Starting with the Keplerians, we imposed a strict 1$\sigma$ Gaussian prior on the orbital period of the inner transiting planet, $P_{\rm b}$, derived from the posterior distribution of the same variable in the transit photometry analysis. Similarly, we imposed a strict Gaussian prior to the time of periastron passage, $t_{\rm p,b}$, inflating the $\sigma$ to account for the uncertainties in the eccentricity of the planet. The period of the outer planet was bound by a uniform prior between [75,150], derived from the minimum period allowed by consecutive TESS photometry and the information derived from the periodogram analysis. Given the inability to derive a period from transit photometry, the time of periastron passage of the outer planet $t_{\rm p,c}$ was bound by a uniform prior in the range [2459678.5, 2459773.5], determined by the preliminary $P_{\rm c}$ from the periodograms. $S_{k}$ and $C_{k}$ for both planets are also bound by uniform priors in the range [-1, 1]. The SOPHIE-HARPS-N offset was allowed to vary only in the [-5,5] \ms interval. The rest of the parameters are left with wide positive (larger than zero) uniform priors.

Regarding the kernel hyperparameters, we applied a strict Gaussian prior to $\theta_{4}$ (the "smoothness" of the fit) centred on 0.5$\pm$0.05, as recommended by \citet{Jeffers2009}. This choice is grounded in the fact that even highly complex active-region distributions average out to just two or three large active regions per rotation.
We set a wide Gaussian prior on the stellar rotation period $\theta_{2}$ derived from the periodogram analysis centred in 48 days with $\sigma$=10 days, as wide as the forest of peaks in the WASP BGLS periodogram. The evolution timescale $\theta_{3}$ is bound by a wide Jeffreys' prior. A Jeffreys' prior is a uniform, uninformed prior that is invariant under reparameterization of the given parameter vector. It is less informative than a uniform prior when the scale and range of the considered parameter is not known, as it corresponds to a uniform probability density in logarithmic frequency. A wide positive (larger than zero) uniform prior is applied to the GP amplitude $\theta_{1}$, and the jitter is only allowed to vary in the interval [0,2] \ms.

\subsection{The Eccentricity of TOI-2134c}
\label{sec:bimecc}

Initial analysis of the radial-velocity data showed a significant trimodality in the distribution of the eccentricity of the outer 95 days-orbit planet, $e_{\rm c}$. After further investigation we found that multiple fully-converged models with different outer planet eccentricities existed. The RVs allowed for eccentricities of TOI-2134c equal to $0.002^{+0.029}_{-0.002}$, $0.45\pm0.05$ and  \eccc$^{\ecccu}_{\ecccd}$. All the models agreed within their uncertainties for most other parameters.
Significantly large eccentricities have been detected before for temperate gas planets (as mentioned in Section \ref{sec:intro}) and stability can be reached within this system, so we could not a priori exclude any of the models. The stellar rotational period derived from the analysis is close to half the period of TOI-2134c. We therefore postulated that an interaction between the fit of the Keplerian model and the stellar activity-induced signal by the GP could be the reason behind the multiple models. While the flexibility of GPs are what makes them valuable tools to model stellar activity, we believe that in this case this flexibility allowed the Keplerian to take different accepted forms, while absorbing any "left-over" signal into the activity model.
To further compare the final likelihoods of the three solutions, we computed the corrected Akaike Information Criterion, AICc, \citep{Sugiura1978} for all converged models:
\begin{equation}
    \mathrm{AICc} =  \mathrm{AIC} + 2 \left(\frac{N_{\mathrm{free}} (1 + N_{\mathrm{free}}  ) }{N_{\mathrm{data}} - N_{\mathrm{free}} +1 }  \right),
\end{equation}
where $N_{\mathrm{free}}$ is the number of free parameters and $N_{\mathrm{data}}$ is the number of data points. The original Akaike Information Criterion, AIC, \citep{Akaike1983} is calculated as
\begin{equation}
    \mathrm{AIC} = -2 \ln \mathbb{L} + 2 N_{\mathrm{free}},
\end{equation}
where ln$\mathbb{L}$ is the logarithmic likelihood maximised after the MCMC analysis. The larger the AICc the less likely the model.
The AICcs of the combined (HARPS-N + SOPHIE) RV data for the low-, medium- and high-eccentricity models were respectively 1224.0 and 1195.7 and 1196.7. As a further check, and to test whether this system would significantly benefit from a simpler analysis, we also computed the Keplerian-only best-fit model to the data. For this analysis we only included the planetary model with a jitter term and no stellar activity or GP component. This last model struggled to converge and its AICc was 1253.2. This analysis led us to strongly disfavour the Keplerian-only model and the circular-orbit model (with AICc difference from the best model larger than 7). However, the AICc values for the medium and the high eccentricity cases were similar enough that no single model was significantly favoured and no significant statistical preference could be reached.\\
\begin{figure}
	\includegraphics[width=\columnwidth]{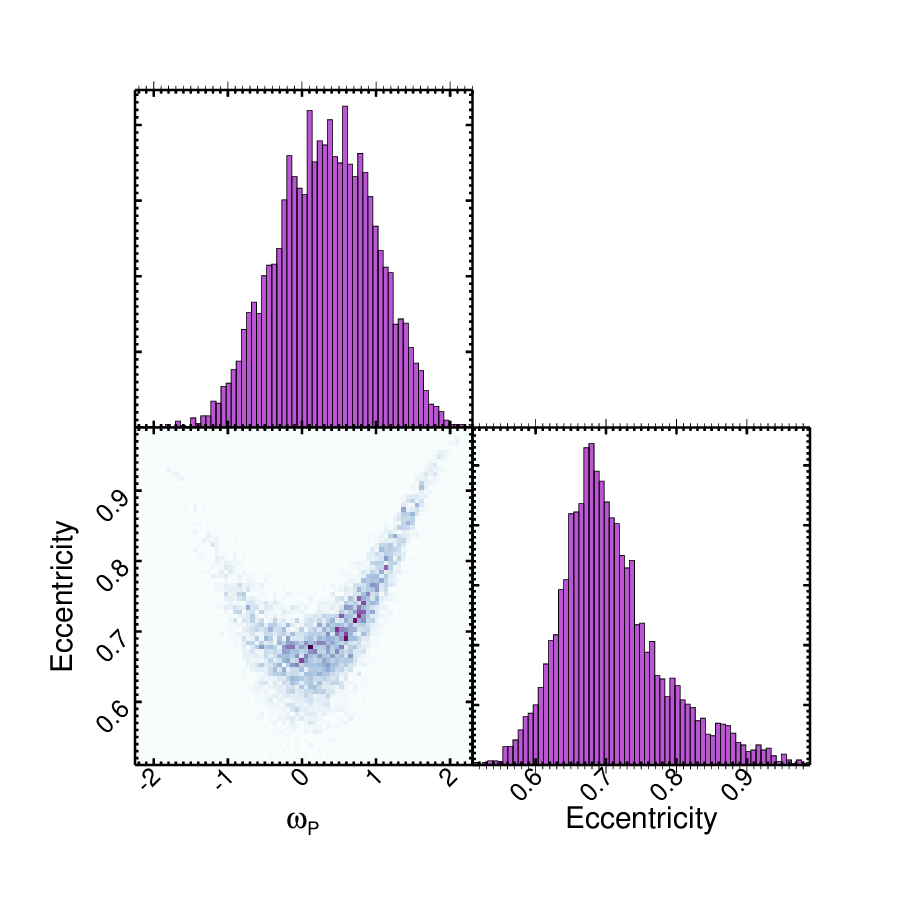}
    \caption{Posterior distribution corner of the eccentricity and the argument of periastron $\omega_{\rm p}$ of the outer planet c derived after MCMC model optimisation on the deep mono-transit present in the TESS data, as explained in Section \ref{sec:bimecc}. Most notable, the eccentricity of TOI-2134c converges to a high $\sim$0.7 value.}
    \label{fig:tr_eo}
\end{figure}

We then turned to the obtained photometric data.
We estimated the orbital period of the singly-transiting planet candidate using only the TESS light curve, following the procedure of \cite{Vanderburg2018}. This method does not take into consideration the results from radial velocity, and derives the planetary period directly from the photometric mono-transit. We extracted the impact parameter $b_{\rm c}$, planet-star radius ratio $R_{\rm c}/R_\star$, and total transit duration of the single transit candidate from the MCMC posteriors from our two-planet transit fit, and solved for the orbital period assuming the stellar parameters reported in this paper and an eccentricity probability distribution from \cite{Kipping2014}. We also imposed the constraint that a second transit was not observed by TESS, which requires the orbital period be longer than about 75 days. We found that the short duration of the transit and minimum period allowed by TESS rule out circular orbits for this planet with periastron passage happening near the time of transit (as expected from geometric arguments), as the RV model comparison also had found.
We then estimated the eccentricity $e_{\rm c}$ and argument of periastron $\omega_{\rm p,c}$ required to reproduce the transit data, assuming the orbital period larger than 75 days. The posterior probability distributions of $e_{\rm c}$ and $\omega_{\rm p,c}$ are shown in Fig. \ref{fig:tr_eo}. The eccentricity is required to be high ($\sim$0.7), and the argument of periastron is broadly to happen near the conjunction of the orbit of the planet.

This eccentricity value derived from transit photometry was then used to constrain the RVs. Given the high-eccentricity preference, we added a Gaussian prior centred in 0.7 with a $\sigma$ of 0.1 to $e_{\rm c}$. In this paper we chose to only report the high-eccentricity RV models for the HARPS-N, SOPHIE and the combined RV dataset consistent with the results from photometry.

\subsection{RV Results}
A summary of the final results of our RV analyses can be found in Table \ref{tab:separategp}. For this MCMC analysis we simultaneously evolved 100 chains for 100,000 iterations each, discarding a burn-in phase of 20,000 steps. We assessed the health and convergence of the chains by computing the Gelman-Rubin statistic and all parameters reached values under the 1.1 convergence cut. As mentioned in the previous section we tested a series of models. For each set of HARPS-N only, SOPHIE only and combined RVs we evolved Keplerian-only models with no stellar activity (which overall struggled to converge or did not converge), forced circular-orbit models, medium-eccentricity models, and finally high-eccentricity models bound with an eccentricity prior derived by the photometry analysis. In this paper we only present the last set.

The HARPS-N only data can constrain the amplitude and period of the inner TOI-2134b better than the SOPHIE data can, but conversely the SOPHIE RVs are able to better identify the signal of the outer planet, especially its period. A combined analysis allows us to more robustly constrain both planets with a single model. Since all three of the Gaussian Process regression models fully converged and reached final values consistently within 1$\sigma$ of each other, we only discuss the results of the combined RV analysis.

The periods of the two planets are well defined. Their RV amplitudes are constrained to 12$\sigma$ for planet b and to 6$\sigma$ for planet c.
The MCMC struggles to constrain the stellar activity evolution timescale $\theta_2$, as expected from the low correlation with activity indicators and the weak overall rotational modulation (see Section \ref{sec:act,rv}). The stellar rotation period is derived to be \Pk$_{\Pkd}^{\Pku}$ days.

\section{Joint Photometry and RV analysis}
\label{sec:joint}

Finally, we also modelled the TESS photometry and the radial-velocity data jointly, to more robustly test whether the high eccentricity model was still favoured. This more complex analysis allowed for simultaneous modelling of the orbital solutions for both planets.
We once again used the code \texttt{MAGPy\_RV}\footnote{This version of MAGPy\_RV is not yet public.}, which for joint photometry analysis includes transit modelling with the python package \texttt{batman} \citep{Kreidberg2015}.

We modelled the RVs similarly to Section \ref{sec:rv}, as two Keplerian signals for the planet candidates with a Quasi-Periodic kernel describing the stellar activity and an offset parameter to match the zero-line of the HARPS-N and the SOPHIE datasets. For the TESS data we described the transits of both planets with six parameters each (period, time of transit, $S_{\rm k}$, $C_{\rm k}$, planet to stellar radius ratio, and orbital inclination).
Our photometric model also included five parameters to describe the host star (its mean density, $q_1$, $q_2$, photometric jitter and offset). In this analysis we are jointly modelling the periods, time of transits, eccentricity and angle of periastron of each planet.

\subsection{Selection of Priors}
We imposed the similar priors on the GP hyperparameters as described in Section \ref{sec:gpprior}: Gaussian priors on the stellar rotational period and the harmonic complexity, uniform priors on amplitude and RV jitter, and a Jeffreys' prior on the evolution timescale.
The RV offset between SOPHIE and HARPS-N data was also similarly bound by a uniform prior between [-5,5]. The period of the inner planet, $P_{\rm b}$, was bound by a Gaussian prior centred on 9.2 days with $\sigma$ of 0.2 days derived from preliminary transit analysis. The time of transit $t_{\rm tr,b}$ was also similarly bound by a Gaussian prior. The period of the outer planet, $P_{\rm c}$, was bound by a uniform prior between [75, 150] days, as it was in the original RV analysis.
The RV amplitude of both planets were as before bound between [0,20] \ms. $S_{\rm k}$ and $C_{\rm k}$ of both planets were only allowed to vary in the interval [-1,1] by definition. For the photometry, the stellar density was bound by a Gaussian prior centred on the derived density in Section \ref{sec:star} with $\sigma$ equal to its uncertainty. We allowed both planet-to-star radius ratios, $R_{\rm b}/R_\star$ and $R_{\rm c}/R_\star$, to only vary between [0,1] (we expect the planets to be smaller then the star), $q_1$ and $q_2$ between [0,1] as per their definition, and we required both inclinations $i$ to be less than 90$^{\circ}$. All other priors were flat uninformative priors.

\begin{table}
\centering
\caption{Results and uncertainties of the planetary parameters for the joint photometry and RV analysis described in Section \ref{sec:joint}}\label{tab:joint}.
\begin{tabular}{ccc}
\hline
\hline
\rule{0pt}{0ex} \vspace{-0.2cm} \\
Parameter & Value\\
\rule{0pt}{0ex} \vspace{-0.2cm} \\
\hline
\rule{0pt}{0ex} \vspace{-0.2cm} \\
    GP Amplitude $\theta_1$ [\ms] & 4.59$_{-1.29}^{+1.38}$  \\ [2 pt]
    GP Timescale $\theta_2$ [days] & 28.01$_{-22.15}^{+21.31}$ \\[2 pt]
    GP Period $\theta_3$ [days] & 53.87$_{-3.02}^{+3.14}$ \\[2 pt]
    GP Smoothness $\theta_4$ & 0.44$_{-0.06}^{+0.08}$ \\[2 pt]
    Jitter [\ms] & 0.85$_{-0.59}^{+0.95}$ \\[2 pt]
    SOPHIE HARPS-N Offset [m/s] & 2.64$_{-0.06}^{+0.09}$ \\[2 pt]

    Orbital period $P_{\rm b}$~[days] & 9.229209$_{-0.000004}^{+0.000006}$\\[2 pt]
    Radius ratio $(R_{\rm b}/R_\star)$ & 0.02$\pm$0.01 \\
    Orbital inclination $i_{\rm b}$~[deg] & 89.91$_{-0.06}^{+0.05}$  \\ [2 pt]
    RV Amplitude $K_{\rm b}$ [\ms] & 3.51$_{-0.41}^{+0.33}$\\[2 pt]
    Eccentricity $e_{\rm b}$ & 0.05$_{-0.03}^{+0.03}$ \\[2 pt]
    Argument of periastron $\omega_{\rm p,b}$ [rad] & -0.75$_{-0.88}^{+0.47}$\\[2 pt]
    Time of periastron $t_{\rm p,b}$ [BJD] & 2459407.82$_{-0.06}^{+0.09}$\\[2 pt]

    Orbital period $P_{\rm c}$~[days] & 94.98$_{-1.02}^{+0.95}$\\[2 pt]
    Radius ratio $(R_{\rm c}/R_\star)$ & 0.09$\pm$0.01 \\
    Orbital inclination $i_{\rm b}$~[deg] & 89.91$_{-0.03}^{+0.02}$ \\[2 pt]
    RV Amplitude $K_{\rm c}$ [\ms] & 9.83$_{-0.89}^{+0.85}$ \\[2 pt]
    Eccentricity $e_{\rm c}$ & 0.62$_{-0.02}^{+0.09}$\\[2 pt]
    Argument of periastron $\omega_{\rm p,c}$ [rad] & 1.41$_{-0.48}^{+0.49}$\\[2 pt]
    Time of periastron $t_{\rm p,c}$ [BJD] & 2459432.39$_{-3.01}^{+3.11}$ \\[2 pt]
\rule{0pt}{0ex} \vspace{-0.2cm} \\
\hline
\end{tabular}	
\end{table}

\begin{table}
\centering
\caption{System Parameters for the \thisstar\ system. The transit and radial-velocity parameters are derived in Sections \ref{sec:transit} and \ref{sec:rv}. Derived parameters are addressed in Section \ref{sec:results} and its subsections alongside the necessary assumptions. \label{tab:big}}
\begin{tabular}{ccc}
\hline
\hline
\rule{0pt}{0ex} \vspace{-0.2cm} \\
Parameter & Value\\
\rule{0pt}{0ex} \vspace{-0.2cm} \\
\hline
\rule{0pt}{0ex} \vspace{-0.2cm} \\
\multicolumn{2}{c}{\emph{Gaussian Process Regression - Modelled Activity Parameters}}\\
\rule{0pt}{0ex} \vspace{-0.3cm} \\
    GP Amplitude $\theta_1$ [\ms] & \Kk$_{\Kkd}^{\Kku}$  \\ [2 pt]
    GP Timescale $\theta_2$ [days] & \Evok$_{\Evokd}^{\Evoku}$ \\[2 pt]
    GP Period $\theta_3$ [days] & \Pk$_{\Pkd}^{\Pku}$ \\[2 pt]
    GP Smoothness $\theta_4$ &\Harmk$_{\Harmkd}^{\Harmku}$ \\[2 pt]
    Jitter [\ms] & \jit$_{\jitd}^{\jitu}$ \\[2 pt]
    SOPHIE HARPS-N Offset [m/s] & \offset$_{\offsetd}^{\offsetu}$ \\[2 pt]
\rule{0pt}{0ex} \vspace{-0.2cm} \\
\hline
\rule{0pt}{0ex} \vspace{-0.2cm} \\
\multicolumn{2}{c}{\textbf{TOI-2134 b}}\\
\multicolumn{2}{c}{\emph{Transit and Radial-Velocity Parameters}}\\
\rule{0pt}{0ex} \vspace{-0.3cm} \\
    Orbital period $P_{\rm b}$~[days] & \perplb$\pm$\uperplb\\
    Time of transit $t_{\rm 0,b}$~[BJD] & \ttransitb$\pm$\uttransitb\\ 
    Radius ratio $(R_{\rm b}/R_\star)$ & \rprstb$\pm$ \urprstb \\
    Orbital inclination $i_{\rm b}$~[deg] & \inclb$\pm$ \uinclb \\ 
    Transit impact parameter $b_{\rm b}$ & \impb$\pm$\uimpb  \\
    Transit duration $\tau_{\rm b}$ [hours] & 2.995$\pm$0.047 \\
    RV Amplitude $K_{\rm b}$ [\ms] & \Kb$_{\Kbd}^{\Kbu}$\\[2 pt]
    Eccentricity $e_{\rm b}$ & \eccb$^{\eccbu}_{\eccbd}$ \\[2 pt]
    Argument of periastron $\omega_{\rm p,b}$ [rad] & \omegab$^{\omegabu}_{\omegabd}$\\[2 pt]
    Time of periastron $t_{\rm p,b}$ [BJD] & \tperib$^{\tperibu}_{\tperibd}$\\
\rule{0pt}{0ex} \vspace{-0.2cm} \\
\hline
\rule{0pt}{0ex} \vspace{-0.2cm} \\
\multicolumn{2}{c}{\emph{Derived Parameters}}\\
\rule{0pt}{0ex} \vspace{-0.3cm} \\
    Radius $R_{\rm b}$ [\rearth] & \rplb$\pm$\urplb \\
    Mass $M_{\rm b}$ [\mearth] & \Mb$^{\Mbu}_{\Mbd}$\\[2 pt]
    Density $\rho_{\rm b}$~[kg m$^{-3}$] &  2607$\pm$516 \\
    Density $\rho_{\rm b}$ [$\rho_{\oplus}$] &  0.47$\pm$0.09 \\
    Scaled semi-major axis ($a_{\rm b}/R_\star$)  &  23.66$\pm$0.52\\
    Semi-major axis $a_{\rm b}$ [AU] &  0.0780$\pm$0.0009 \\
    Incident Flux $F_{\rm inc,b}$~[$F_{\rm inc,\oplus}$] &  32$\pm$2\\
    Equilibrium temperature $T_{\rm eq,b}$~[K] &  666$\pm$8 \\
\rule{0pt}{0ex} \vspace{-0.2cm} \\
\hline
\rule{0pt}{0ex} \vspace{-0.2cm} \\
\multicolumn{2}{c}{\textbf{TOI-2134 c}}\\
\multicolumn{2}{c}{\emph{Transit and Radial-Velocity Parameters}}\\
\rule{0pt}{0ex} \vspace{-0.3cm} \\
    Orbital period $P_{\rm c}$~[days] & \Pc$_{\Pcd}^{\Pcu}$\\[2 pt]
    Time of transit $t_{\rm 0,c}$~[BJD] & \ttransitc$\pm$\uttransitc\\ 
    Radius ratio $(R_{\rm c}/R_\star)$ & \rprstc$\pm$ \urprstc\\
    Transit impact parameter $b_{\rm c}$ & \impc$\pm$ \uimpc \\
    Transit duration $\tau_{,c}$ [hours] & 5.267$\pm$0.028  \\
    RV Amplitude $K_{\rm c}$ [\ms] & \Kc$_{\Kcd}^{\Kcu}$ \\[2 pt]
    Eccentricity $e_{\rm c}$ & \eccc$_{\ecccd}^{\ecccu}$\\[2 pt]
    Argument of periastron $\omega_{\rm p,c}$ [rad] & \omegac$_{\omegacd}^{\omegacu}$\\[2 pt]
    Time of periastron $t_{\rm p,c}$ [BJD] & \tperic$_{\tpericd}^{\tpericu}$ \\
\rule{0pt}{0ex} \vspace{-0.2cm} \\
\hline
\rule{0pt}{0ex} \vspace{-0.2cm} \\
\multicolumn{2}{c}{\emph{Derived Parameters}}\\
\rule{0pt}{0ex} \vspace{-0.3cm} \\
    Radius $R_{\rm c}$~[\rearth] & \rplc$\pm$ \urplc\\
    Mass $M_{\rm c}$~[\mearth] & \Mc$_{\Mcd}^{\Mcu}$ \\[2 pt]
    Density $\rho_{\rm c}$~[kg m$^{-3}$] &  599$\pm$152 \\
    Density $\rho_{\rm c}$ [$\rho_{\oplus}$] &  0.11$\pm$0.03 \\
    Scaled semi-major axis ($a_{\rm c}/R_\star$)  &  112$\pm$2 \\
    Semi-major axis $a_{\rm c}$ [AU] &  0.371$\pm$0.004 \\
    Incident Flux $F_{\rm inc,c}$~[$F_{\rm inc,\oplus}$] &  1.4$\pm$0.1 \\
    Equilibrium temperature $T_{\rm eq,c}$~[K] & 306$\pm$4 \\
\rule{0pt}{0ex} \vspace{-0.2cm} \\
\hline
\end{tabular}	
\end{table}

\begin{figure*}
	\includegraphics[width=17cm]{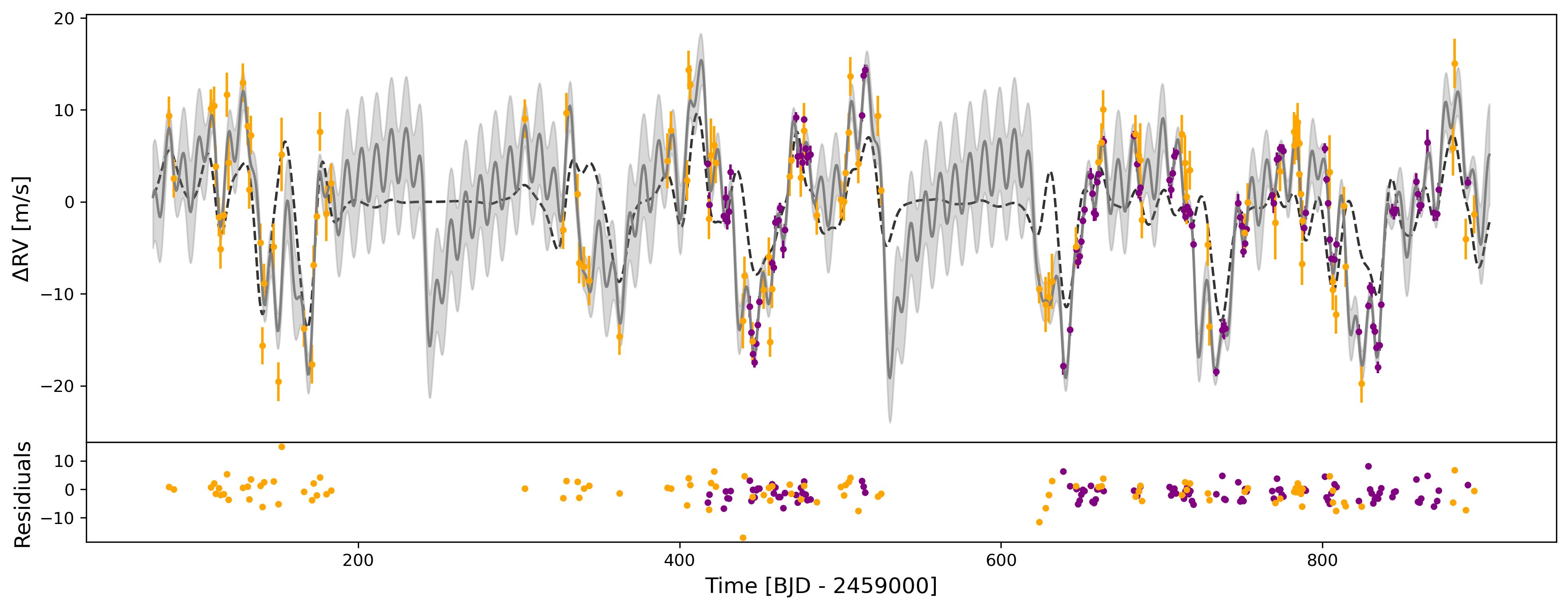}
    \caption{Combined SOPHIE (orange) and HARPS-N (purple) radial-velocity data plotted with errorbars (HARPS-N errorbars are too small to be clearly visible). The complete model, which includes two Keplerians and the predicted activity, is plotted in grey, with its uncertainties as the gray shaded area. The dashed black line represents the GP activity prediction only. On the bottom, the residuals between the data (in the corresponding colour) and the complete model are plotted.}
    \label{fig:gprv}
\end{figure*}
\subsection{Joint Analysis Results}

We simultaneously evolved 100 chains for 100,000 iterations each, discarding once again a burn-in phase of 20,000 steps and we tested for convergence with the Gelman-Rubin statistic. The results of our combined analysis are listed in Table \ref{tab:joint}. All parameters agree within 1$\sigma$ uncertainty with the results from the previous less complex transit and RV analyses, shown in Tables \ref{tab:phot} and \ref{tab:separategp}.
These results once again confirmed the high-eccentricity model for the outer planet TOI-2134c.

Overall, we were able fully recover both planet candidates and their periods. Their RV amplitudes were constrained to 10$\sigma$ for the inner planet and 11$\sigma$ for the outer one. The joint photometry and RV analysis is minorly less effective in the retrieval of the RV signal of inner planet than the radial-velocity data on their own, but it performed better for TOI-2134c. Once again, the stellar activity evolution timescale is not very well-constrained. The stellar rotational period was here derived to be slightly longer (54.27$_{-3.23}^{+3.27}$) but it is still consistent with the previous analysis. Both planet radius ratios were fully retrieved to 2 and 9$\sigma$ for TOI-2134 b and c respectively.

\begin{figure}
    \centering
    \includegraphics[width=\columnwidth]{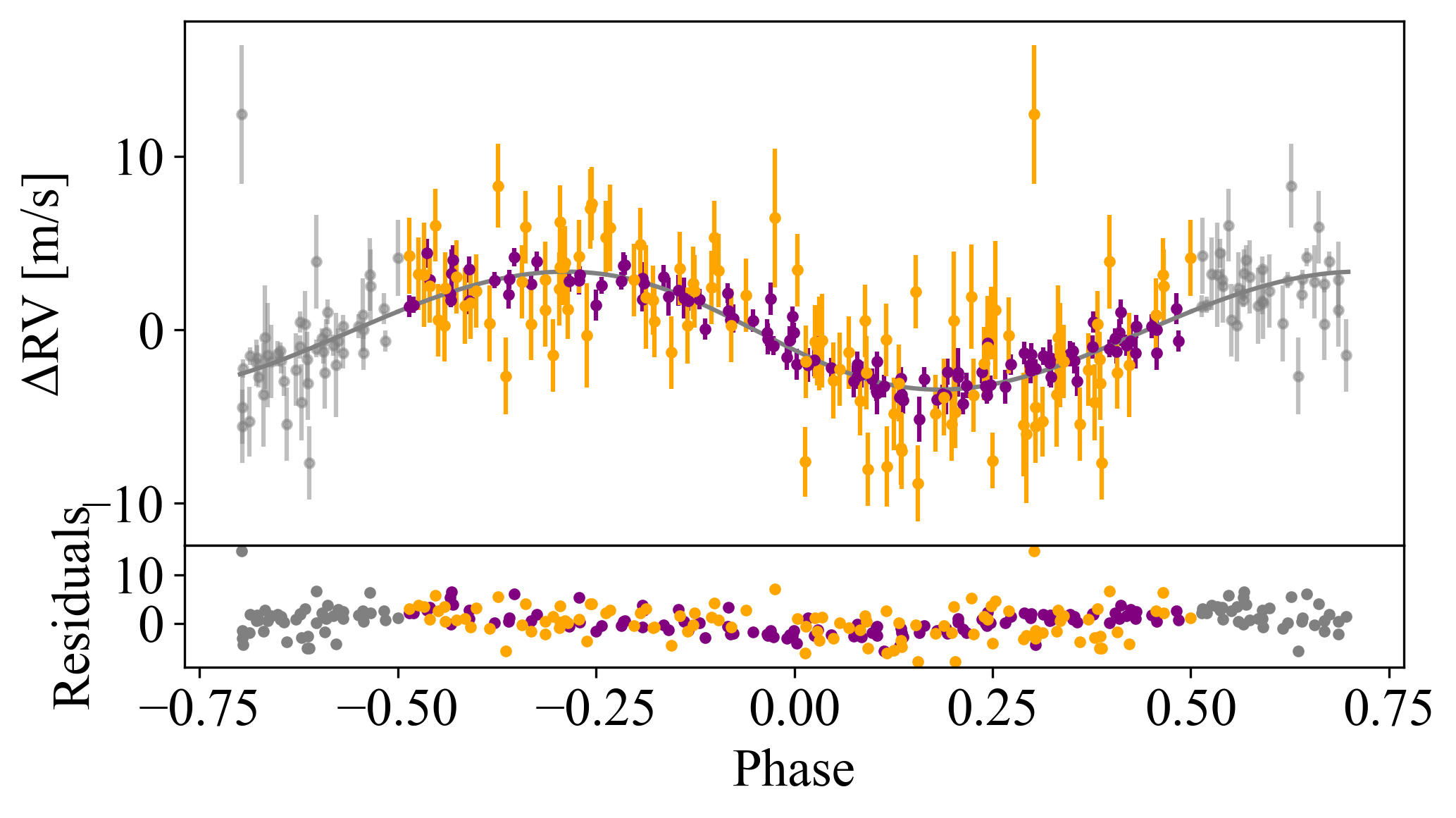}
    \includegraphics[width=\columnwidth]{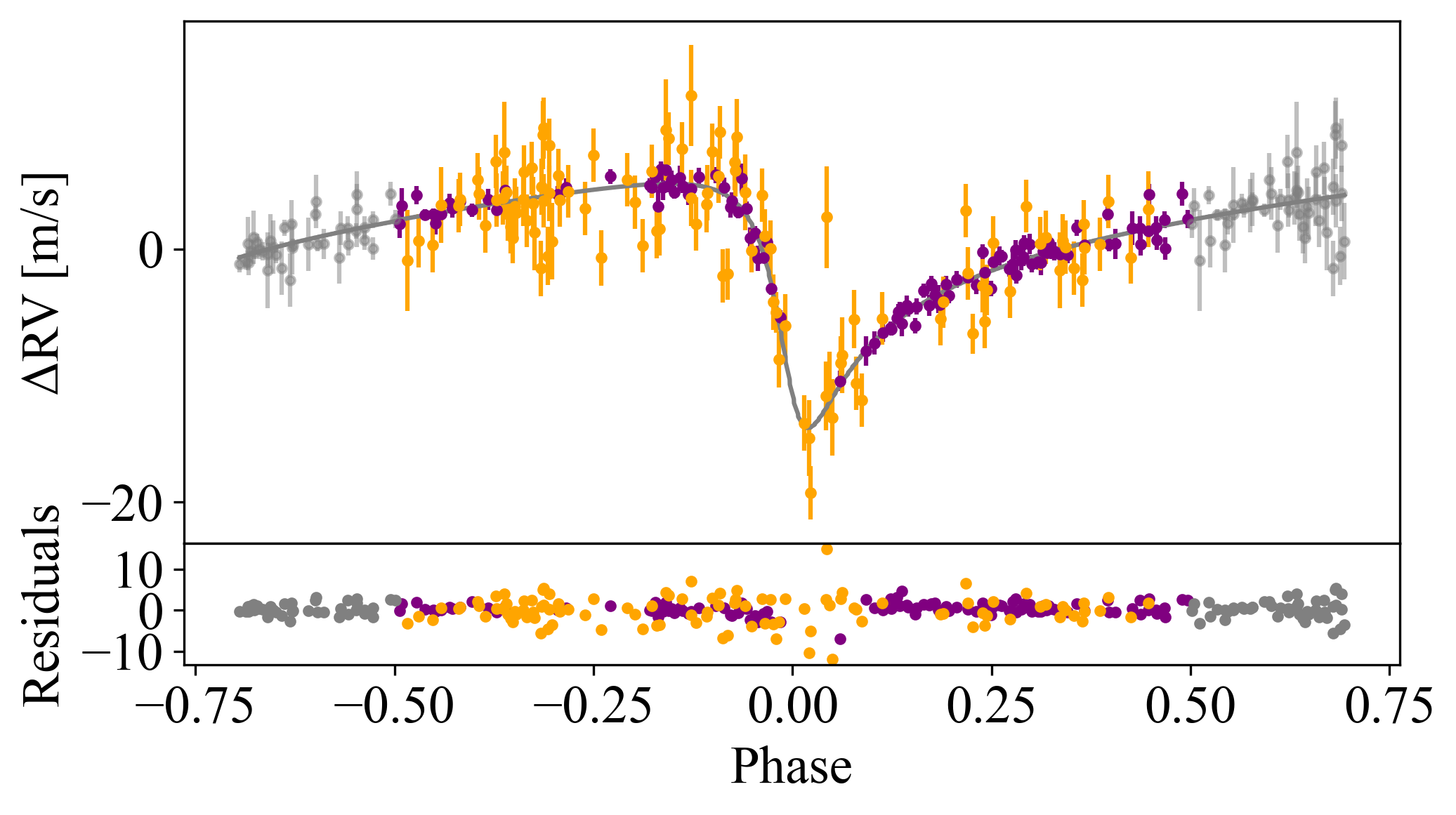}
    \caption{Phase folded activity model-subtracted plots for the inner (top) and outer (bottom) planets. In orange are the SOPHIE RVs and in purple the HARPS-N ones with respective errorbars (some HARPS-N errorbars may be too small to be visible). The Keplerian model is plotted as a gray line, with the residuals shown on the bottom. The phase has also been extended on both sides.}
    \label{fig:phases}
\end{figure}

\section{Results and Discussion}
\label{sec:results}

The results of the joint photometry and RV analysis fully agree within their 1$\sigma$ uncertainties with the results from the separate transit and RV analyses. While the joint method successfully retrieved and characterised both planet candidates, from here on, we chose to use the results from the less complex, separated analyses undertaken in Sections \ref{sec:transit} and \ref{sec:rv}. All the final results are compiled in Table \ref{tab:big}. In Fig. \ref{fig:gprv} we plot the combined SOPHIE and HARPS-N dataset alongside the complete best-fit model in grey, as well as the GP-predicted activity as a black dashed line. Fig. \ref{fig:phases} shows the phase folded, best-fit Keplerian orbital models, after subtracting the stellar activity-induced signal modelled by the GP, and their residuals.\\

\begin{figure*}
	\includegraphics[width=16cm]{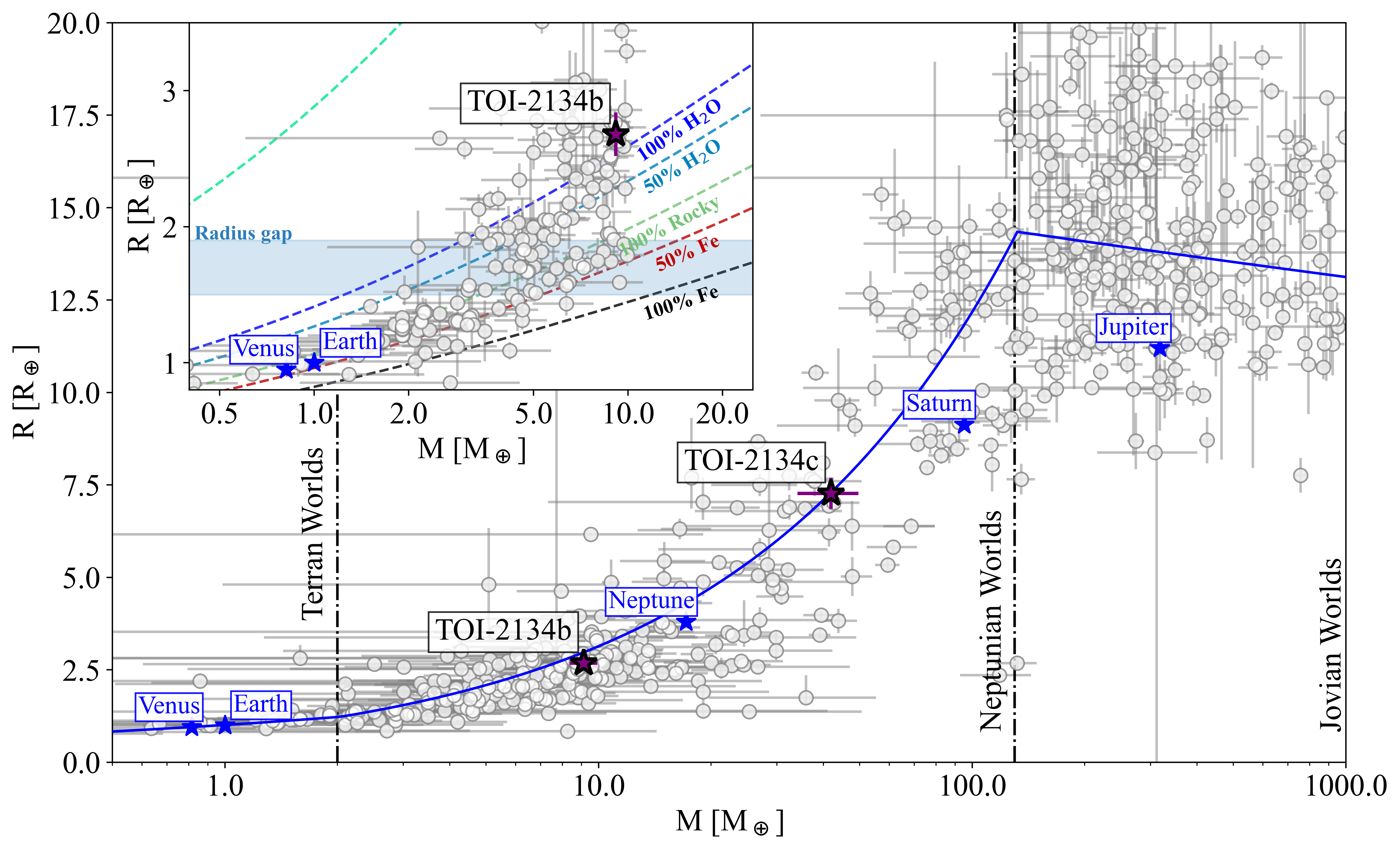}
    \caption{Mass-radius diagram with zoom-in for sub-Neptunian planets. The data are taken from the EU Exoplanet catalogue: \url{http://exoplanet.eu/catalog/} on 17 Feb 2023. The full blue line shows the mass-radius relation developed by \protect\cite{Chen2017}, with its categorisation of Terran (M<2M$_{\earth})$, Neptunian (2M$_{\earth}$<M<0.4M\textsubscript{J}) and Jovian worlds (M>0.4M\textsubscript{J}). The zoomed-in plot includes composition lines taken from \protect\cite{Zeng2016}, and the Radius Valley band. Solar system planets are included for scale.}
    \label{fig:MR}
\end{figure*}

As a result of our investigation, we establish the presence of an inner planet TOI-2134b, and an outer planet TOI-2134c. All derived planetary characteristics are listed in Table \ref{tab:big}. Fig. \ref{fig:MR} shows the two planets in a mass-radius diagram.

We computed for the inner planet TOI-2134b a mass $M_{\rm b}$ of \mbox{\Mb$_{\Mbd}^{\Mbu}$ M$_{\earth}$} and a radius of \mbox{2.69$\pm$0.16 R$_{\earth}$}, for an orbital period of \mbox{9.2292005$\pm$0.0000063 days}. Combining mass and radius yielded a bulk density of \mbox{0.47$\pm$0.09 $\rho_{\earth}$}. In the mass-radius diagram TOI-2134b falls in a parameter space significantly degenerate in composition. Planet b could be a 100\% water-planet \citep{Zeng2016}. At the same time it could also have a rocky core, a water (or other heavy volatile elements) layer and a low-mass H/He envelope. Overall, it is not possible to distinguish the composition of planet b without additional information. For more information about the atmospheric characteristics of TOI-2134b see \cite{Zhang2023}.\\
The outer planet TOI-2134c has mass $M_{\rm c}$ of \mbox{\Mc$_{\Mcd}^{\Mcu}$ M$_{\earth}$} and a radius of \mbox{7.27$\pm$0.42 R$_{\earth}$}, for a period of \mbox{\Pc$_{\Pcd}^{\Pcu}$ days}. The derived mass from the RVs and radius from photometry are well-matched and further justify the association of the mono-transit and the detected radial-velocity signal. The presence of a third planet with similar mass to TOI-2134c that could instead explain the transit would have been detected in the radial-velocity.
The bulk density of TOI-2134c is calculated to be \mbox{0.11$\pm$0.03 $\rho_{\earth}$ }(similar to the density of Saturn). It can therefore be considered a long-orbit mini-Saturn.
Given its derived period, we also went back to the other photometric data and computed when transits would have occurred. The derived transit times are plotted in Fig. \ref{fig:WASP_data} as black dashed lines, and their uncertainty windows as gray shaded regions. TOI-2134c transited 5 times over the 3 years of WASP coverage, but none of those transits was originally detected. The possible explanation for this is twofold. On one hand, WASP is a ground instrument and therefore only observes during dark hours; given the transit duration of $\sim$5 hours, the event could have easily been missed. At the same time, the precision of the WASP data fluctuates significantly and a 0.01 flux deficit (as it is for TOI-2134c) is often too shallow for WASP to reliably detect.

\subsection{System Orbital Stability}
As a preliminary test of the stability of the system given the high eccentricity of TOI-2134c, we calculated the radius of the Hill Sphere \citep{Hamilton1992} of the outer planet and compared it to the closest approach distance between the two planets. If the orbit of the inner TOI-2134b at any point falls within the Hill Sphere of TOI-2134c, we expect the two bodies to gravitationally interact enough to de-stabilise their orbits.
If a body of mass $m$ is orbiting a larger body of mass $M$ at semi-major axis $a$ with an eccentricity $e$, the Hill Radius $R_{\rm Hill}$ of the smaller body can be approximated to be
\begin{equation}
    R_{\rm Hill} \approx a(1-e)\sqrt[3]{{\frac{m}{3M}}}.
\label{eq:Rhill}
\end{equation}
For planet c we computed a $R_{\rm Hill, c}$ of 0.0047$\pm$0.0008 AU. The closest approach between the outer and inner planets is 0.048$\pm$0.026 AU. Therefore, the orbit of planet b at no point intersects with the Hill Sphere of TOI-2134c.\\
\begin{figure}
    \centering
    \includegraphics[width=\columnwidth]{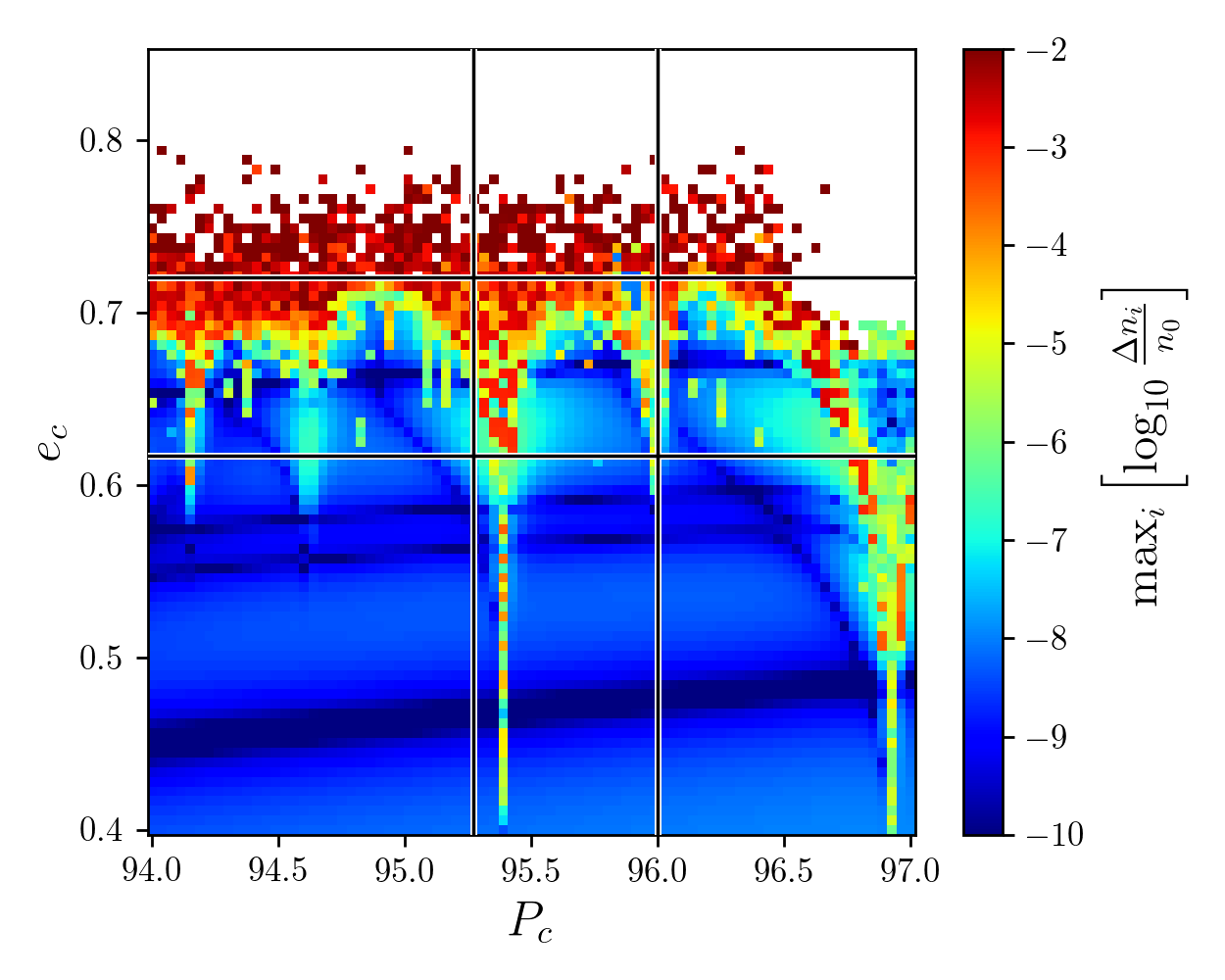}
    \caption{Chaos map for the outer planet TOI-2134c. The period $P_{\rm c}$ and eccentricity $e_{\rm c}$ are explored on a 81x81 grid of different system configurations. After numerical integrations the NAFF indicator is computed and plotted as a colorscale. Blue regions correspond to weakly chaotic, therefore more stable, planetary systems, while red areas refer to strongly chaotic systems, and hence more unstable. The best-fit system position in this space together with its 1$\sigma$ uncertainties indicate that both stable and unstable solutions are compatible with our high-eccentricity fit.}
    \label{fig:stability}
\end{figure}
To further assess the stability of the system under the high eccentricity $e_c$ model, we also computed the chaos map in the neighbourhood of the best-fit solution to the high-eccentricity model, shown in Fig. \ref{fig:stability}. We created a grid of 81x81 system configurations that vary between each other based on period $P_{\rm c}$ and eccentricity $e_{\rm c}$. All other parameters were fixed to their values derived from the MCMC best-fit estimation. Each system defineed a unique set of initial conditions that was then used for 50 kyr numerical integrations with \texttt{REBOUND}\footnote{REBOUND is an open-source software package dedicated to N-body integrations: \url{http://rebound.readthedocs.org}} \citep{Rein2012}  with the 15$^{\rm th}$ order adaptive time-step integrator \texttt{IAS15} \citep{Rein2015}. We also included in our analysis the correction from general relativity implemented in the \texttt{REBOUND} extension \texttt{REBOUNDx}\footnote{Available at \url{https://reboundx.readthedocs.io}} by \citet{Tamayo2020}. After the simulations, we computed the Numerical Analysis of Fundamental Frequencies \citep[NAFF:][]{Laskar1990,Laskar1993}. The NAFF indicator informs about the amount of chaos in a planetary orbit by precisely estimating its main frequency via a technique called frequency analysis \citep{Laskar1988}. The main frequency of a planetary orbit corresponds to the mean-motion, which does not drift over time in non-chaotic dynamics, but does drift if the system is chaotic. Therefore, we apply frequency analysis on the two halves of each simulation, and for each planetary orbit, to estimate the amount of drift in the mean-motions. Weakly chaotic (hence stable) orbits should only show small differences in mean motions between the two integration halves. In this work we consider as the NAFF of the system the logarithmic maximum value of this drift, defined as
\begin{equation}
    {\rm NAFF} = {\rm max}_{i} \left[ {\rm log}_{10} \frac{\Delta n_i}{n_0} \right],
\end{equation}
in which $i$ refers to the chosen planet, $\Delta n_i$ is the difference in the mean-motion of planet $i$ between its estimation on the first and second halves of the integrations, and $n_0$ is the initial mean motion of that planet $i$.
In Fig. \ref{fig:stability}, blue regions have lower NAFF, and are weakly chaotic. Red regions correspond to systems that undergo strong chaos, and likely lead to rapid instability\footnote{We refer to \citet{Stalport2022} for details on the link between NAFF and orbital stability.}. White regions refer to those systems which had an escape or a close encounter between two bodies, and for which the simulation was stopped. We also overplot the area of 1$\sigma$ limit uncertainties on the estimates of $P_c$ and $e_c$. Inside the subsequent square, we find that both chaotic and regular systems can exist. In other words, the high eccentricity model is not incompatible with the system stability.

\subsection{Planetary Incident Flux and Equilibrium Temperature}
\label{sec:temp}
The incident flux of a planet $F_{\rm inc}$ is computed from stellar luminosity $L_{\star}$ and planetary semi-major axis $a$ with the following formula:
\begin{equation}
    F_{\rm inc} = \frac{L_{\star}}{4\pi a^2} = \frac{4\pi R_{\star}^{2} \, \sigma_{\rm SB} \, T_{\rm eff}^{4}}{4\pi a^2},
\label{eq:Finc}
\end{equation}
where $T_{\rm eff}$ and $R_{\star}$ are the stellar effective temperature and radius and $\sigma_{\rm SB}$ is the Stefan-Boltzmann constant.
We can express this same incident flux in Earth units as:
\begin{equation}
    \frac{F_{\rm inc}}{F_{\rm inc,\oplus}} = \left( \frac{T_{\rm eff}}{T_{\odot}}\right)^{4} \left( \frac{R_{\star}}{R_{\odot}}\right)^{2} \left( \frac{1}{a}\right)^{2},
\label{eq:Finc2}
\end{equation}
in which $T_{\odot}$ and $R_{\odot}$ are the solar effective temperature and radius and $a$ is expressed in AU.
Given semi-major axes $a_{b}$ and $a_{c}$ of 0.0780$\pm$0.0009 and 0.371$\pm$0.004 AU respectively, we computed incident fluxes of 33$\pm$2 and 1.4$\pm$0.1 $F_{\rm inc, \oplus}$ for planet b and c.\\
\begin{figure}
    \centering
    \includegraphics[width=\columnwidth]{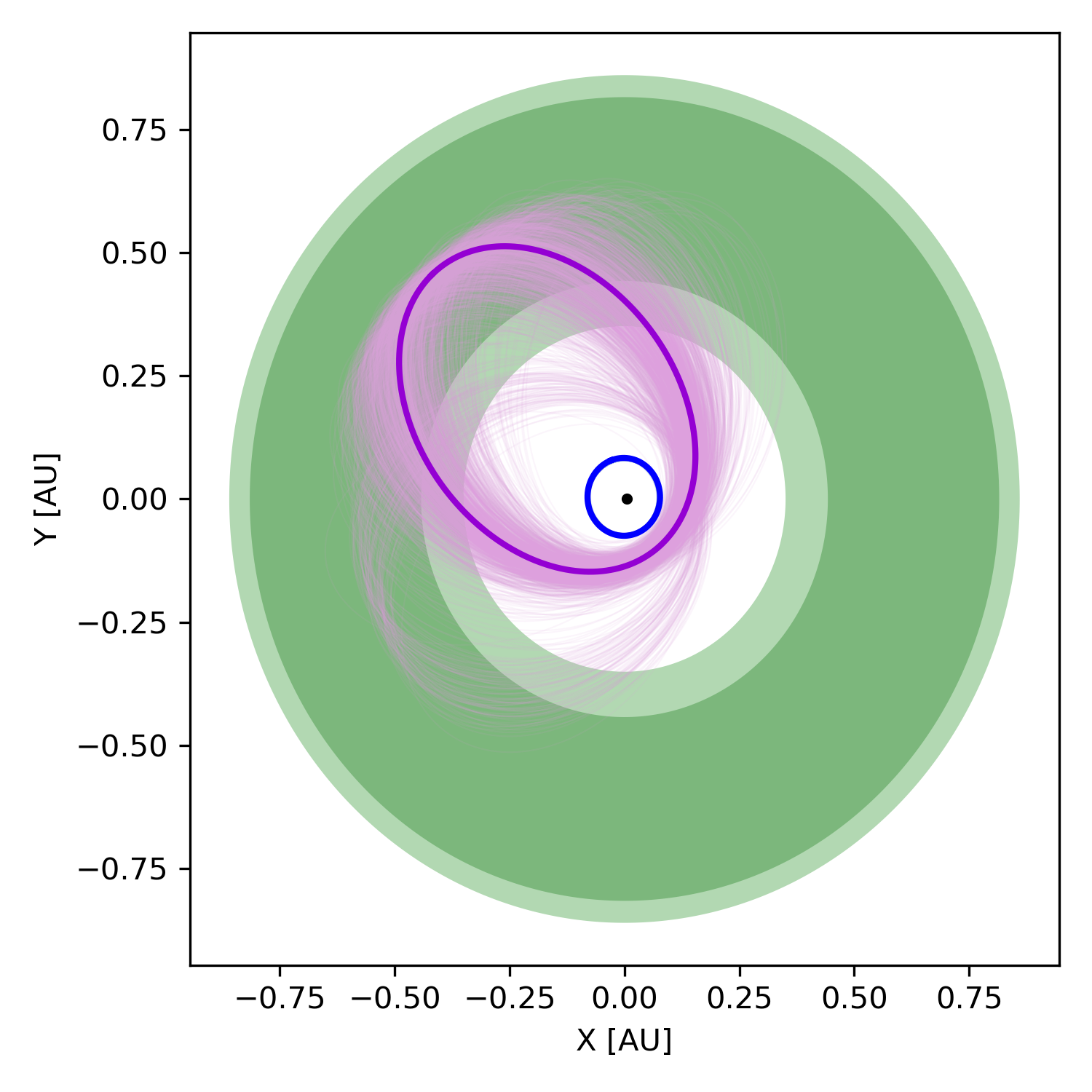}
    \caption{Depiction of the configuration of the TOI-2134 system. We include the inner planet with a circular orbit of $9.2292004\pm0.0000063$ days in blue, and the outer planet with an eccentric ($e_{\rm c}$ = $\eccc^{\ecccu}_{\ecccd}$) orbit of \mbox{$\Pc^{\Pcu}_{\Pcd}$ days} in purple. Their uncertainties are depicted as lighter orbits. The habitable zone boundaries are indicated as green shaded regions: the empirical HZ is plotted in lighter green, while the narrow HZ is overplotted in darker green. The boundaries are computed as described in Section \ref{sec:temp} based on results from \citet{Kopparapu2014}.}
    \label{fig:orbit}
\end{figure}

The planets' equilibrium temperatures $T_{\rm eq}$ can be derived as
\begin{equation}
    T_{\rm eq} = T_{\rm eff} \sqrt{\frac{R_{\star}}{2a}} [f(1-A_{B})]^{1/4},
\label{eq:Teq}
\end{equation}
where $A_{\rm B}$ is the Bond albedo of the considered planet and $f$ represents the effectiveness of atmospheric circulation. Assuming isotropic re-emission and a uniform equilibrium temperature over the entire planet (therefore $f = 1$), an upper limit on $T_{\rm eq}$ can be derived from Equation \ref{eq:Teq} by setting \mbox{$A_{\rm B}=0$}.
We, therefore, calculated the upper limit of the equilibrium temperature of planet b to be 666$\pm$8 K, and of planet c to be 305$\pm$4 K.

From this analysis the upper limit of the equilibrium temperature of the sub-Saturn object would be compatible with liquid water. Planet c is a gas giant, but could host potentially temperate rocky moons.
However, the orbit of TOI-2134c is highly eccentric and the distance of the planet from the star changes significantly during its orbit, as shown in purple in Fig. \ref{fig:orbit}. The boundaries of the habitable zone (HZ) of the system, $r_{\rm HZ, \star}$, can be derived from the solar luminosity $L_{\odot}$ and the stellar luminosity as:
\begin{equation}
     \frac{L_{\odot}}{r_{\rm HZ, \odot}} = \frac{L_{\star}}{r_{\rm HZ, \star}},
\label{eq:HZ}
\end{equation}
where $r_{\rm HZ, \odot}$ is the radius of the boundaries of the solar HZ. The boundaries in this paper were determined following the two models for narrow and empirical habitable zones described in \citep{Kopparapu2014}. The narrow HZ is bound by an inner Runaway Greenhouse limit and an outer Maximum Greenhouse limit. The boundaries of the empirical HZ are defined by the Recent Venus and Early Mars limits. The narrow and empirical HZs for the TOI-2134 system are shown in Fig. \ref{fig:orbit} respectively in dark and light green. As Fig. \ref{fig:orbit} clearly shows, TOI-2134c only spends less than half of its orbit within the HZ boundaries. In fact, we also computed the incident flux and upper limit of the equilibrium temperature planet c at periastron to be \mbox{13$\pm$4 $F_{\rm inc, \oplus}$} and 533$\pm$8 K.

\begin{table}
\centering

\caption{List of times of transits of TOI-2134c between the detected mono-transit and the end of 2025. The uncertainty on the dates computed as shown in Section \ref{sec:followups,tr}. The transit that should be observed by TESS in Sector 80 is highlighted in bold.
\label{tab:list}}

\begin{tabular}{ccc}

\rule{0pt}{0ex} \\
\hline
\hline
\rule{0pt}{0ex} \\
BJD & UT Date (yyyy-mm-dd) & UT Time (hh:mm:ss) \\
\rule{0pt}{0ex} \\
\hline
\rule{0pt}{0ex} \\
2459814.5$\pm$0.3 & 2022-08-22 & 23:20:35\\
2459910.0$\pm$0.6 & 2022-11-26 & 11:25:12 \\
2460005.5$\pm$0.9 & 2023-03-01 & 23:29:53\\
2460101.0$\pm$1.2 & 2023-06-05 & 11:34:34\\
2460196.5$\pm$1.5 & 2023-09-08 & 23:39:11\\
2460292.0$\pm$1.8 & 2023-12-13 & 11:43:52\\
2460387.4$\pm$2.1 & 2024-03-17 & 23:48:29\\
\textbf{2460483.0$\pm$2.4} & \textbf{2024-06-21} & \textbf{11:53:10}\\
2460578.5$\pm$2.7 & 2024-09-24 & 23:57:50\\
2460674.0$\pm$3.0 & 2024-12-29 & 12:02:28\\
2460769.5$\pm$3.3 & 2025-04-04 & 00:07:08\\
2460865.0$\pm$3.6 & 2025-07-08 & 12:11:46\\
2460960.5$\pm$3.9 & 2025-10-12 & 00:16:26\\
\rule{0pt}{0ex} \\
\hline
\rule{0pt}{0ex} \\

\end{tabular}
\end{table}
\subsection{Suggested Follow-Up Observations}
\label{sec:followups}

\subsubsection{Long-term RV Observations and Transit Detection for TOI-2134c}
\label{sec:followups,tr}
This system would foremost benefit from long-term radial-velocity observations to better constrain the period and eccentricity of the outer planet. Both HARPS-N and SOPHIE plan on continuing observing the star sporadically.
A second photometric observing campaign aimed at detecting another transit of the outer planet candidate would also be valuable. In the current mission plan, TESS will re-observe TOI-2134 in Sectors 74, 79 and 80 in 2024. A transit of planet c should occur in Sector 80 (late June to early July 2024).
Given the brightness of TOI-2134 and the larger radius ratio between planet c and its host star, transits of the outer planet can also be observed with ground-based telescopes. Another firm detection of a transit would re-confirm its period and further inform the eccentricity model choice. We include a list of the times of transit between the original detection and the end of 2025 in Table \ref{tab:list}. The uncertainties on the times of transit $\sigma_{\rm tr}$ increase with increasing number of "missed" transits as:
\begin{equation}
    \sigma_{\rm tr} = \sqrt{(n\sigma_{P})^2 + \sigma_{t_{0}}} \approx n\sigma_{P},
\label{eq:Ttrerr}
\end{equation}
in which $n$ is the epoch since the observed transit, and $\sigma_{P}$ and $\sigma_{t_{0}}$ are the uncertainties on respectively the period of the planet and its observed transit time.

\subsubsection{Rossiter-McLaughlin Analysis}
Given the presence of both the inner mini-Neptune and the outer temperate sub-Saturn, (once the period of the outer planet is better defined with follow up RV observations or a second transit detection), TOI-2134 and its planets are scientifically valuable targets for follow-up Rossiter-McLaughlin (RM: \citealt{Rossiter1924, McLaughlin1924}) analysis to determine the spin-orbit alignment of the system. The RM amplitude $K_{\rm RM}$ can be computed as
 \begin{equation}
     K_{\rm RM} = 52.8 {\rm m s^{-1}}  \frac{v {\rm sin}(i)}{5 {\rm km s^{-1}}} \left(\frac{R_{\rm pl}}{R_{\rm J}}\right)^2 \left(\frac{R_{\star}}{R_{\odot}}\right)^{-2} , 
 \end{equation}
in which $R_{\rm pl}$ and $R_{\star}$ are the radius of the considered transiting planet and the radius of the star. Instead of using a maximum limit for $v {\rm sin}(i)$, we recomputed it starting from the derived stellar rotational period to be 0.78$\pm$0.09 \kms. Since both TOI-2134b and c transit, we computed the minimum expected RM amplitude for both: $K_{\rm RM,b}= 0.98\pm0.17$ \ms and $K_{\rm RM,c}=7.2\pm1.2$ \ms. Although the longer transit duration can be an obstacle, RM observations of temperate gas giants as TOI-2134c are valuable to further our understanding of planet migration. A significant fraction of hot giants are shown to have orbits that are misaligned with the rotational axis of their star \citep{Winn2010, Albrecht2012}. The origin of such misalignment is still unclear, but a leading hypothesis is that high-eccentricity migration tilts the orbit of the planet away from its initial plane via dynamical interactions (e.g. \citealt{Ford2008,Fabrycky2007,Petrovich2015}). Unlike hot giants, it is significantly more challenging to form temperate gas planets via high-eccentricity migration \citep{Dong2013}, and it is even less likely in the case of this system due to the presence of an inner small planet. Therefore, if high-eccentricity migration is in fact the driving factor behind the misalignment, the majority of temperate giants should have orbits aligned to spin of their star.
However, given their lower transit probabilities, there are only few RM observations of temperate giants. Whether the aim is the whole transit or just observing the ingress or egress in a shorter summer night, the temperate sub-Saturn planet c has a large peak-to-peak amplitude (\mbox{$7.2\pm1.3$ \ms}) that makes it easily observable. With a more firmly constrained eccentricity model, TOI-2134c will be a great candidate for RM follow-up.

\subsubsection{Transmission Spectroscopy}
We also discussed the suitability of TOI-2134b and c for follow-up atmospheric characterisation via transmission spectroscopy. \cite{Kempton2018} developed an analytic metric to estimate the expected SNR of transmission-spectroscopy observations based on the strength of the spectral features and the brightness of the star, the Transmission Spectroscopy Metric, or $TSM$. It can be computed as:
\begin{equation}
    TSM = \epsilon \cdot \frac{R_{\rm pl}^{3}T_{\rm eq}}{M_{\rm pl} R_{\star}^{2}} \cdot 10^{-m_{\rm J}/5}, 
\end{equation}
in which $R_{\rm pl}$ and $M_{\rm pl}$ are the radius and mass of the considered planet in Earth radii and masses, $R_{\star}$ is the stellar radius in solar radii, $T_{\rm eq}$ is the equilibrium temperature of the planet computed at zero albedo and full day-night heat redistribution (as in Section \ref{sec:temp}), and $m_{\rm J}$ is the apparent magnitude of the host star in the J-band. The term $\epsilon$ is a normalisation factor to give one-to-one scaling to the JWST/NIRISS 10-hour simulated observations described in \cite{Louie2018}. This scaling factor also absorbs the unit conversion factors so that the parameters can be in natural units. The term $\epsilon$ changes depending on the radius of the planet, and is equal to 1.26 for TOI-2134b, and 1.15 for TOI-2134c. We computed a $TSM_{\rm b}=172\pm42$ and a $TSM_{\rm c}=243\pm54$. The $TSM$s of both planets are therefore considered well above the suggested cut-offs for their size bin. It is however important to note that the $TSM$ was developed for targeted JWST effort and therefore it is not optimised for stars with $m_{\rm J} < 9$ mag, as brighter stars require the bright readout mode and have substantially lower duty cycles. Given its brightness, TOI-2134 currently is observable without saturation by the JWST with NIRCam in its bright mode, with similar observational strategies as the ones successfully proposed by Dr. Hu for 55 Cancri e (Program ID: 1952) and by Dr. Deming for HD 189733b (Program ID: 1633). Moreover, higher efficiency read modes for JWST observations are being investigated \citep{Batalha2018} and future dedicated missions such as Ariel, and the ground-based ELTs are suitable for brighter targets such as TOI-2134 \citep{Danielski2022, Houlle2021}.

\section{Summary and Conclusions}

In this work we presented the photometric light curves of five TESS sectors and of three years of WASP monitoring, alongside 219 high-precision radial-velocity measurements obtained with HARPS-N and SOPHIE of the star TOI-2134. We characterised the star with multiple independent techniques and we studied its periodograms to better understand its stellar activity signals. We then performed a transit photometry analysis on the photometric data and a Gaussian Process regression analysis on the radial-velocity data to constrain the radii and masses of the planets in the system. To test the statistical strength of the derived model we also completed a joint analysis of the photometric and the RV data. The resulting planetary parameters fully agree within 1$\sigma$ uncertainties with the results of the previous investigations. We therefore selected to focus on the results of the less complex, separate analyses for our discussion.
As a result, we reached the following conclusions:

\begin{itemize}
    
    \item
    We characterise the new multi-transiting inner planet \mbox{TOI-2134b} in a 9.2292004$\pm$0.0000063 day orbit with \mbox{$M_{\rm b}=\Mb^{\Mbu}_{\Mbd}$ M$_{\oplus}$} (12$\sigma$ detection) and \mbox{$R_{\rm b}=2.69\pm0.16$ R$_{\oplus}$}. Its bulk density \mbox{($\rho_{b}$ = 0.47$\pm$0.09 $\rho_{\oplus}$)} identifies the planet as either a water-world or a mini-Neptune with a rocky core and a low-mass H/He envelope. We computed the upper limit of the equilibrium temperature of the planet to be 666$\pm$8 K.

    \vspace{0.2cm}
    \item
    We also constrain a second mono-transiting planet TOI-2134c with  \mbox{$M_{\rm c}=\Mc^{\Mcu}_{\Mcd}$M$_{\oplus}$} (5$\sigma$ detection) and \mbox{$R_{\rm c}=7.27\pm0.42$ R$_{\oplus}$} in a \mbox{$\Pc^{\Pcu}_{\Pcd}$ days} orbit, with an upper limit of the equilibrium temperature of \mbox{306$\pm$4 K}. Its bulk density ($\rho_{c}$ = 0.11$\pm$0.03 $\rho_{\oplus}$) is similar to Saturn's.

    \vspace{0.2cm}

    \item
    After GP regression, we find three possible orbital architectures for the outer TOI-2134c that model the radial-velocity data, one with low eccentricity (0.0002$^{+0.0025}_{-0.0002}$), one with medium eccentricity (0.45$\pm$0.05), and one with high eccentricity ($\eccc^{\ecccu}_{\ecccd}$). While we were able to disfavour the circular orbit case, the AICc values of the latter two solutions are comparable, therefore statistically there is no preference. We noted that in all models the rotational period of the star is half the orbital period of the outer TOI-2134c. We postulated that fitting interactions between the Keplerian model for the planet, and the activity-induced signal that the GP is extrapolating are the reason behind the multiple fully-converged solutions. The flexibility of the GP allows the Keplerian to take different accepted forms while the GP model absorbs the residual signal and attributes it to stellar activity. As described in Section \ref{sec:bimecc}, further analysis of the photometry data showed that, given the derived orbital period for planet c, its transit duration time was too short to allow circular orbits. In fact, the mono-transit in the TESS data strongly prefers the high eccentricity case. To further strengthen our results, we also undertook joint modelling of the photometric and the RV data. This investigation yielded a single converged state with an $e_{\rm c}$=0.61$^{+0.08}_{-0.03}$. In this paper we therefore chose to present the high eccentricity model of the separate, less complex RV only analysis and to use it for all further analysis. We also tested the stability of the system given these results and reached the conclusion that the high-eccentricity model is not incompatible with a stable system.

    \vspace{0.2cm}
    
    \item
    Since the mass-radius parameter space planet TOI-2134c resides in is not well populated and in order to better constrain its period and eccentricity, we recommend further RV observations and a second photometric observing campaign to detect another transit.
    To further characterise the architecture of the system we also recommend Rossiter-McLaughlin follow-up observations. We compute the expected RM amplitude of the temperate sub-Saturn TOI-2134c as \mbox{$7.2\pm1.2$ \ms}, making it accessible to ground-based instruments.

    \vspace{0.2cm}
    
    \item
    We also compute the Transmission Spectroscopy Metric of both planets of the system for possible follow-up atmospheric characterisation via transmission spectroscopy. Although the projected SNRs place the planets well above the recommended cuts, TOI-2134 is close to the bright limits of most instruments on JWST, and is currently only observable with NIRCam in its bright mode. Future missions such as Ariel or ground-based transition spectroscopy will be suited for brighter target such as TOI-2134.

\end{itemize}

\section*{Affiliations}
\noindent
{\it
$^{1}$Department of Astrophysics, University of Exeter, Stocker Rd, Exeter, EX4 4QL, UK\\
$^{2}$Institut d’Astrophysique de Paris, UMR7095 CNRS, Universit\'e Pierre \& Marie Curie, 98bis Boulevard Arago, 75014 Paris, France \\
$^{3}$Observatoire de Haute-Provence, CNRS, Universit\'e d'Aix-Marseille, 04870 Saint-Michel-l'Observatoire, France\\
$^{4}$Kavli Institute for Astrophysics and Space Research, Massachusetts Institute of Technology, 77 Massachusetts Avenue, Cambridge, MA 02139, USA\\
$^{5}$Department of Physics and Astronomy, The University of North Carolina at Chapel Hill, Chapel Hill, NC 27599, USA\\
$^{6}$Department of Astrophysics, University of Birmingham, Edgbaston, Birmingham, B15 2TT, UK\\
$^{7}$DTU Space,  National Space Institute, Technical University of Denmark, Elektrovej 328, DK-2800 Kgs. Lyngby, Denmark\\
$^{8}$Center for Astrophysics | Harvard \& Smithsonian, 60 Garden Street, Cambridge, MA 02138, USA\\
$^{9}$Trottier Institute for Research on Exoplanets (\emph{iREx}))\\
$^{10}$D\'epartement de Physique, Universit\'e de Montr\'eal, 1375 Avenue Th \'er\`ese-Lavoie-Roux, Montreal, QC, H2V 0B3, Canada\\
$^{11}$Astrophysics Group, Keele University, Staffordshire, ST5 5BG, UK\\
$^{12}$Department of Physics, University of Warwick, Gibbet Hill Road, Coventry CV4 7AL, UK\\
$^{13}$Space Sciences, Technologies and Astrophysics Research (STAR) Institute, Universit\'e de Li\`ege, All\'ee du 6 Ao\^ut 19C, B-4000 Li\`ege, Belgium\\
$^{14}$Observatoire de Gene\'eve, Universit\'e de Gene\'eve, Chemin de Pegasi, 51, CH-1290 Versoix, Switzerland\\
$^{15}$University of Southern Queensland, Centre for Astrophysics, West Street, Toowoomba, QLD 4350 Australia\\
$^{16}$Laboratoire d'Astrophysique de Marseille, Universit\'e de Provence, UMR6110 CNRS, 38 rue F. Joliot Curie, 13388 Marseille cedex 13, France\\
$^{17}$Space Telescope Science Institute, 3700 San Martin Drive, Baltimore, MD, 21218, USA\\
$^{18}$Aix Marseille Univ, CNRS, CNES, LAM, Marseille, France\\
$^{19}$Royal Astronomical Society, Burlington House, Piccadilly, London W1J 0BQ, UK\\
$^{20}$SUPA, School of Physics \& Astronomy, University of St Andrews, North Haugh, St Andrews, KY169SS, UK\\
$^{21}$Centre for Exoplanet Science, University of St Andrews, North Haugh, St Andrews, KY169SS, UK\\
$^{22}$Univ. Grenoble Alpes, CNRS, IPAG, 38000 Grenoble, France\\
$^{23}$SUPA, Institute for Astronomy, Royal Observatory, University of Edinburgh, Blackford Hill, Edinburgh, EH9 3HJ, UK\\
$^{24}$Centre for Exoplanet Science, University of Edinburgh, Edinburgh, EH9 3FD, UK\\
$^{25}$Instituto de Astrof{\'\i}sica e Ci\^encias do Espa\c{c}o, Universidade do Porto, CAUP, Rua das Estrelas, 4150-762 Porto, Portugal\\
$^{26}$Department of Astrophysical Sciences, Princeton University, Princeton, NJ 08544, USA\\
$^{27}$INAF - Osservatorio Astrofisico di Torino, Strada Osservatorio, 20 I-10025 Pino Torinese (TO), Italy\\
$^{28}$Department of Earth, Atmospheric and Planetary Science, Massachusetts Institute of Technology, 77 Massachusetts Avenue, Cambridge, MA 02139, USA\\
$^{29}$Department of Aeronautics and Astronautics, MIT, 77 Massachusetts Avenue, Cambridge, MA 02139, USA\\
$^{30}$Laborat\'{o}rio Nacional de Astrof\'{i}sica, Rua Estados Unidos 154, 37504-364, Itajub\'{a} - MG, Brazil\\
$^{31}$SETI Institute, Mountain View, CA 94043, USA/NASA Ames Research Center, Moffett Field, CA 94035, USA\\
$^{32}$Fundaci\'on Galileo Galilei - INAF (Telescopio Nazionale Galileo), Rambla J. A. F. Perez 7, E-38712 Bre\~na Baja (La Palma), Canary Islands, Spain\\
$^{33}$Instituto de Astrof\'{\i}sica de Canarias, C/V\'{\i}a L\'actea s/n, E-38205 La Laguna (Tenerife), Canary Islands, Spain\\
$^{34}$Departamento de Astrof\'{\i}sica, Univ. de La Laguna, Av. del Astrof\'{\i}sico Francisco S\'anchez s/n, E-38205 La Laguna (Tenerife), Canary Islands, Spain\\
$^{35}$Department of Physics and Astronomy, University of New Mexico, 210 Yale Blvd NE, Albuquerque, NM 87106, USA\\
$^{36}$Dipartimento di Fisica e Astronomia "Galileo Galilei" - Università degli Studi di Padova, Vicolo dell'Osservatorio 3, 35122, Padova, Italy\\
$^{37}$INAF - Osservatorio Astronomico di Padova, Vicolo dell'Osservatorio 5, Padova, 35122, Italy\\
$^{38}$INAF - Osservatorio Astronomico di Palermo, Piazza del Parlamento 1 90134, Palermo, Italy\\
$^{39}$ Centre for Astrophysics, University of Southern Queensland, West Street, Toowoomba, AU\\
$^{40}$ Sub-department of Astrophysics, University of Oxford, Keble Rd, OX13RH, Oxford, UK\\
$^{41}$Observatoire des Baronnies Proven\c{c}ales, 05150 Moydans, France\\
$^{42}$Astrobiology Research Unit, Universit\'e de Li\`ege, 19C All\'ee du 6 Ao\^ut, 4000 Li\`ege, Belgium\\
$^{43}$Kotizarovci Observatory, Sarsoni 90, 51216 Viskovo, Croatia\\
$^{44}$American Association of Variable Star Observers, 185 Alewife Brook Parkway, Cambridge, MA 02138, USA\\
$^{45}$Oukaimeden Observatory, High Energy Physics and Astrophysics Laboratory, Cadi Ayyad University, Marrakech, Morocco\\
$^{46}$Instituto de Astrof\'isica de Andaluc\'ia (IAA-CSIC), Glorieta de la Astronom\'ia s/n, 18008 Granada, Spain\\
$^{47}$Grand Pra Observatory, 1984 Les Haudères, Switzerland\\
$^{48}$Observatori Astron\'omic Albany\'a, Girona, Spain\\
$^{49}$Departamento de Astronom\'{\i}a y Astrof\'{\i}sica, Universidad de Valencia, E-46100 Burjassot, Valencia, Spain\\
$^{50}$Observatorio Astron\'omico, Universidad de Valencia, E-46980 Paterna, Valencia, Spain}

\section*{Acknowledgements}

FR is funded by the University of Exeter's College of Engineering, Maths and Physical Sciences, UK.

The HARPS-N project was funded by the Prodex Program of the Swiss Space Office (SSO), the Harvard University Origin of Life Initiative (HUOLI), the Scottish Universities Physics Alliance (SUPA), the University of Geneva, the Smithsonian Astrophysical Observatory (SAO), the Italian National Astrophysical Institute (INAF), University of St. Andrews, Queen’s University Belfast, and University of Edinburgh.

This work has been supported by the National Aeronautics and Space Administration under grant No. NNX17AB59G, issued through the Exoplanets Research Program.

This work has been carried out within the framework of the NCCR PlanetS supported by the Swiss National Science Foundation under grants 51NF40$\_$182901 and 51NF40$\_$205606.

This project has received funding from the European Research Council (ERC) under the European Union’s Horizon 2020 research and innovation programme (grant agreement SCORE No 851555).

ACC and TGW acknowledge support from STFC consolidated grant numbers ST/R000824/1 and ST/V000861/1, and UKSA grant number ST/R003203/1.

FPE and CLO would like to acknowledge the Swiss National Science Foundation (SNSF) for supporting research with HARPS-N through the SNSF grants nr.140649, 152721, 166227 and 184618. The HARPS-N Instrument Project was partially funded through the Swiss ESA-PRODEX Programme.

Funding for the TESS mission is provided by NASA's Science Mission Directorate.KAC acknowledges support from the TESS mission via subaward s3449 from MIT.

This research has made use of the Exoplanet Follow-up Observation Program website, which is operated by the California Institute of Technology, under contract with the National Aeronautics and Space Administration under the Exoplanet Exploration Program.

This paper includes data collected by the TESS mission that are publicly available from the Mikulski Archive for Space Telescopes (MAST).

We thank the Observatoire de Haute-Provence (CNRS) staff for its support in collecting SOPHIE data. This work was supported by the ''Programme National de Plan\'etologie'' (PNP) of CNRS/INSU and CNES.

The postdoctoral fellowship of KB is funded by F.R.S.-FNRS grant T.0109.20 and by the Francqui Foundation.

Resources supporting this work were provided by the NASA High-End Computing (HEC) Program through the NASA Advanced Supercomputing (NAS) Division at Ames Research Center for the production of the SPOC data products.

We acknowledge the use of public TESS data from pipelines at the TESS Science Office and at the TESS Science Processing Operations Center.

DD acknowledges support from the TESS Guest Investigator Program grants 80NSSC21K0108 and 80NSSC22K0185, and from the NASA Exoplanet Research Program grant 18-2XRP18$\_$2-0136.

This work makes use of observations from the LCOGT network. Part of the LCOGT telescope time was granted by NOIRLab through the Mid-Scale Innovations Program (MSIP). MSIP is funded by NSF.

RDH is funded by the UK Science and Technology Facilities Council (STFC)'s Ernest Rutherford Fellowship (grant number ST/V004735/1).

SD is funded by the UK Science and Technology Facilities Council (grant number ST/V004735/1).

BSL is funded by a UK Science and Technology Facilities Council (STFC) studentship (ST/V506679/1).

The postdoctoral fellowship of KB is funded by F.R.S.-FNRS grant T.0109.20 and by the Francqui Foundation.

XD and TF acknowledge funding by the French National Research Agency in the framework of the Investissements d'Avenir program (ANR-15-IDEX-02), through the funding of the "Origin of Life" project of the Grenoble-Alpes University.

\section*{Data Availability}


The observational data presented in this publication are openly available. The TESS data is available at: \url{https://mast.stsci.edu/portal/Mashup/Clients/Mast/Portal.html}. The mentioned NEOSSat data can be found at: \url{https://open.canada.ca/data/en/dataset/9ae3e718-8b6d-40b7-8aa4-858f00e84b30}. The LOCGT data can be found at: \url{https://exofop.ipac.caltech.edu/tess/target.php?id=75878355}.
The WASP data cannot be found online, but can be made available on request. The HARPS-N and SOPHIE radial-velocity data, alongside their mentioned activity proxies are included as supplementary material.



\bibliographystyle{mnras}
\bibliography{mybib} 





\bsp	
\label{lastpage}
\end{document}